\documentclass[12pt]{article}
\usepackage{amssymb,latexsym}
\newtheorem{defi}{Definition}
\newtheorem{coro}{Corollary}
\newtheorem{prop}{Proposition} 
\def\beq{\begin{equation}} 
\def\eeq{\end{equation}}
\def\bea{\begin{eqnarray}} 
\def\eea{\end{eqnarray}}
\def\beas{\begin{eqnarray*}} 
\def\eeas{\end{eqnarray*}} 
\def\nn{\nonumber}
\def\q{\quad}
\def\H{\hat H}
 
\def\Z{\mathbb{Z}} 
\def\N{\mathbb{N}} 
\def\R{\mathbb{R}} 
\def\e{\varepsilon}
\def\O{\Omega}
\def\P{{\cal P}}

\def\mybox{\setbox1=\vbox{\hrule height 4ptwidth 4pt} \hfill\llap{\box1}} 
\begin{document}
\begin{center} 
{\Large \bf 
Jacobson generators, Fock representations\\
and statistics of $sl(n+1)$}\\[5mm] 
{\bf T.D.\ Palev}\footnote{Permanent address~:
Institute for Nuclear Research and Nuclear Energy, 
Boul.\ Tsarigradsko Chaussee 72, 1784 Sofia, Bulgaria;
E-mail~: tpalev@inrne.bas.bg}\\ 
International Centre for Theoretical Physics, 
34100 Trieste, Italy\\[2mm] 
{\bf J.\ Van der Jeugt}\footnote{E-mail~:
Joris.VanderJeugt@rug.ac.be.}\\ 
Department of Applied Mathematics and Computer Science, 
University of Ghent,
Krijgslaan 281-S9, B-9000 Gent, Belgium. 
\end{center}

\vskip 10mm
{\em 
\noindent Corresponding author~: J.\ Van der Jeugt. 
Tel~: ++32 9 2644812. Fax~: ++32 9 2644995. 
E-mail~: Joris.VanderJeugt@rug.ac.be
\vskip 3mm
\noindent Running title~: Quantum statistics of $sl(n+1)$
}
\vskip 10mm
\begin{abstract}
The properties of
$A$-statistics, related to the class of simple Lie algebras
$sl(n+1)$, $n\in \Z_+ $ (Palev, T.D.: Preprint JINR E17-10550
(1977); hep-th/9705032), are further investigated. The
description of each $sl(n+1)$ is carried out via generators
and their relations (see eq.~(\ref{e2-5})), 
first introduced by Jacobson.  
The related Fock spaces $W_p$, $p\in \N$, are finite-dimensional
irreducible $sl(n+1)$-modules.  The Pauli principle of the
underlying statistics is formulated.  In addition the paper
contains the following new results: ($a$) The $A$-statistics are
interpreted as exclusion statistics; ($b$) Within each $W_p$
operators $B(p)_1^\pm,\ldots, B(p)_n^\pm$, proportional to the
Jacobson generators, are introduced.  It is proved that in an
appropriate topology (Definition 2) $\displaystyle\lim_{p\to
\infty} B(p)_i^\pm =B_i^\pm$, where $B_i^\pm$ are Bose creation
and annihilation operators; ($c$) It is shown that the local
statistics of the degenerated hard-core Bose models and of the
related Heisenberg spin models is $p=1$ $A$-statistics.
\end{abstract} 
\vskip 2mm

\section{Introduction}

During the last two decades quantum statistics became a field
of increasing interest among field theorists and condensed matter
theorists.  Various new statistics were suggested, leading to
generalizations or deviations from some of the first principles
in quantum physics, such as the Heisenberg commutation relations,
the Pauli exclusion principle and the commutativity of 
space-time.

The literature on the subject is vast, especially in the part
related to quantum groups~\cite{dr87,ji85,fa89,ma88,wo87}.  
In a paper entitled ``Twisted
Second Quantization"~\cite{pu89} Pusz and Woronowicz introduced multimode
deformed Bose creation and annihilation operators (CAOs),
covariant under the action of the quantum group $U_q[sl(n)]$ (for
$n$ pairs of them). Another deformation with commuting modes of
CAOs was proposed in~\cite{ma89}; the link between them was established
in~\cite{ku90}.  A third deformation, which for one mode of CAOs was
known for many years~\cite{co72}, 
the so called quon algebra~\cite{gr91}, was
defined as an associative algebra, subject to relations
$
a_i^-a_j^+ -q a_j^+a_i^-=\delta_{ij}.
$
This generalization (note that no relations among only
creation operators or among only annihilation operators are
required) was in the origin of a model proposed for a
verification of small violations of Bose-Fermi statistics in
quantum field theory (QFT)~\cite{gr90}.  The quon statistics, which in
the classification of Doplicher, Haag and Roberts~\cite{do71} belongs to
the class of ``infinite statistics", was studied by several
authors~\cite{go89} from different points of view  
(see~\cite{gr99} for further discussions and references).

Recently string theory was also involved in discussions on
quantum statistics, the latter related to its prediction that 
Heisenberg's uncertainty principle has to be corrected at
distances of order of the Plank length $k_P=10^{-32}$~cm.
Consequently there emerges an absolute minimum uncertainty in the
measuring of any length~\cite{ve86}. These predictions motivated several
authors to search for model independent arguments, leading to
the same conclusions as string theory does (we refer to~\cite{ga95} for
a survey on the subject). In particular it has been shown that
the above results can be reproduced on a purely kinematical level
with appropriate deformations of the Heisenberg commutation
relations~\cite{ma93,ke94,hi96,ad98}, i.e., of canonical quantum statistics. 
In all such cases the coordinates do not commute at small distances, a
result which is consistent with the spirit of non-commutative
geometry~\cite{co86}.

Turning to condensed matter physics we refer to
anyons, ``particles" with fractional statistics (FS) in two-dimensional
(2D) systems~\cite{le77}. The theoretical studies of this and other
noncanonical statistics were strongly  pushed  forward after the
discovery of the fractional quantum Hall effect (FQHE) in
two-dimensional electron gases~\cite{ts82}. 
Its theoretical explanation led Laughlin~\cite{la83} 
to the conclusion that there exist quasiparticles carrying
fractional electric charges. The statistics of these particles
(we write ``particles" for the elementary excitations, the
``quasiparticles", when no confusion can arise) 
also turned out to be fractional statistics~\cite{ha84}.

A further breakthrough in the area of quantum statistics was
marked with the 1991 paper of Haldane~\cite{ha91}, who proposed a
generalized version of the Pauli exclusion principle.
For only one kind of identical particles this new statistics, now
called (fractional) exclusion statistics (ES), asserts that the
change $\Delta d$ in the dimension $d$ of the single-particle
Hilbert space is defined via the relation
\beq
{\Delta d}=-g \cdot {\Delta N}. \label{e1-1}
\eeq
Here $\Delta N$ is {\it an allowed increase} of the number of
particles. The constant $g$ is called an exclusion statistics
parameter.

In~\cite{wu94} Wu proposed an ``integral form" compatible
with Haldane's definition~(\ref{e1-1})~:
\beq
d(N)=n-g(N-1). \label{e1-2}
\eeq
In~(\ref{e1-2}) $N-1$ is the number of particles already accommodated in
the system, $d(N)$ is the dimension of the single-particle space,
namely the number of the orbitals, where an additional $N^{th}$
particle can be ``loaded", holding the distribution of the initial
$N-1$ particles fixed; $n=d(1)$ is the number of orbitals
available for the first particle. Eq.~(\ref{e1-2}) holds for {\it all
admissible values} of $N$.

Contrary to Bose or Fermi statistics the orbitals of an ES may
not be filled independently of each other. An essential
difference of ES in comparison to fractional statistics is that
in general the former is defined for any space dimension.
Initially ES was defined for finite-dimensional single-particle
Hilbert spaces~\cite{ha91}. The generalization to infinite-dimensional
cases is due to Murthy and Shankar~\cite{mu94}. 
In~\cite{wu94} Wu extended the
meaning of species. His definition allows different species
indices to refer to particles of the same kind but with different
quantum numbers.

In~\cite{ha91} Haldane has shown that when applied to FQHE, 
ES leads to the
same predictions as FS does (see also~\cite{jo94}). 
The validity of ES was tested on several other examples~: spinon
excitations in a spin-${1\over 2}$ quantum antiferromagnetic chain
(with nearest neighbor-exchange or with inverse-square exchange
between all sites)~\cite{ha91}; anyon gas and anyons in a strong
magnetic field (confined to the first Landau 
level)~\cite{wu94, mu94, li94}; 
particles in 1D Luttinger liquid~\cite{mu94}; Calogero-Sutherland
models~\cite{ha94}.

The discovery of the Yangian $Y(SU_N)$-symmetry of $SU_N$ quantum
chains with inverse-square exchange~\cite{ha92} (generalizations of the
$S={1\over 2}$ Haldane-Shastry spin chains~\cite{ha88}) casted a bridge
between exclusion statistics and the $(SU_N)_1$ Wess-Zumino-Witten
(rational) conformal field theories, providing a new,
alternative, description of these theories (see~\cite{ha93} for a broader
review on the subject). Instead of primary chiral fields, the
fundamental fields in this picture are ``free" $SU_N$-spinon fields,
namely fields which interact only via statistical interaction~\cite{sc94}.
The statistics is encoded in the generalized ``commutation" relations
between the creation and the annihilation operators, namely between the
Fourier modes of the fields.  As a result the single particle Fock
states are occupied in such a way that
the corresponding statistics is an exclusion statistics 
(we refer to~\cite{sc97}
for further details, remarks and additional references).

Despite of the fact that ES is defined for any space dimension,
so far it was applied and tested only within 1D and 2D models.
In the present paper we show that particular kinds of ES, called
$A$-statistics, can exists in spaces with any dimension.

Our approach to quantum statistics is strongly influenced by the
ideas of Wigner, outlined in his 1950's work ``Do the equations of
motion determine the quantum mechanical commutation 
relations?"~\cite{wi50}.  
This was the first paper where it was clearly indicated
that the canonical quantum statistics may, in principle, be
generalized in a logically consistent way. Wigner demonstrated
this on the example of a one-dimensional oscillator with a
Hamiltonian ($m=\omega=\hbar=1$) $H={1\over 2}(p^2 + q^2)$.
Abandoning the requirement $[p,q]=-i$, Wigner was searching for
all operators $q$ and $p$, such that the ``classical" equations of
motion ${\dot p}=-q$, ${\dot q}=p$ are identical with the
Heisenberg equations ${\dot p}=-i[p,H]$, ${\dot q}=-i[q,H]$.
Apart from the canonical solution he found infinitely many other
solutions.  Let ${\sqrt 2}B_1^\pm=q \mp i p$. It turns out~\cite{pa81}
that all these different operators satisfy one and the same triple
relation, namely~(\ref{e1-3}) below with $i=j=k=1,$ (see
the end of this Introduction for the notation)~:
\beq
[\{B_i^\xi,B_j^\eta\},B_k^\varepsilon]=
\delta_{ik}(\varepsilon-\xi)B_j^\eta +
\delta_{jk}(\varepsilon-\eta)B_i^\xi, \quad i,j,k \in \N, \quad
\xi,\eta,\varepsilon= \pm,~\pm 1.
\label{e1-3}
\eeq
The operators $B_i^\pm$, $i=1,2,\ldots$ are para-Bose (pB)
operators, discovered by Green~\cite{gr53} three years later as a
possible generalization of statistics of tensor fields in
QFT. Thus the infinitely many different solutions found by Wigner
were in fact the Fock representations of one pair of para-Bose
operators.

It is known that the linear span of all operators
$B_i^\xi$, $\{B_j^\eta, B_k^\varepsilon\}$ is a Lie 
superalgebra~\cite{om76} 
isomorphic to the orthosymplectic Lie superalgebra
$osp(1/2n)$ for $i,j,k=1,\ldots,n$ and $ \xi,\eta,\varepsilon=
\pm$~\cite{ga80}. The para-Bose operators constitute a basis in the odd
subspace of this superalgebra and generate it. Consequently the
representation theory of $n$ pairs of pB operators is completely
equivalent to the representation theory of $osp(1/2n)$. Hence
Wigner found all Fock representations of $osp(1/2)$ long before
Lie superalgebras (and supersymmetry) became of interest
in physics and even before they were introduced in mathematics.

Similarly, any $n$ pairs of para-Fermi CAOs
$F_1^\pm,F_2^\pm,\ldots,F_n^\pm$~\cite{gr53}, defined by relations
\beq
[[F_i^\xi,F_j^\eta],F_k^\varepsilon]= {1\over 2}
\delta_{jk}(\varepsilon-\eta)^2 F_i^\xi -{1\over
2}\delta_{ik}(\varepsilon-\xi)^2 F_j^\eta, \quad i,j,k \in \N,\quad
\xi,\eta,\varepsilon= \pm,~\pm 1, 
\label{e1-4}
\eeq
generate the Lie algebra $so(2n+1)$~\cite{ka62, ry63}. The key observation
here is that both $so(2n+1)$ and $osp(1/2n)$ belong to class
$B$ of the basic Lie superalgebras in the classification of Kac~\cite{ka79}.
Hence parastatistics (and in particular Bose and Fermi
statistics) appear as particular Fock representations of Lie
superalgebras from one and the same class, the Lie
superalgebras of class $B$. In this sense Green's parastatistics 
could be called $B$-(para)statistics.

The clarification of the mathematical structure, hidden in
parastatistics, provides a natural background for further
searches of new quantum statistics. One such possibility is to
consider deformations of parastatistics, namely deformations of
$so(2n+1)$ and $osp(1/2n)$ in the sense of quantum groups. We
refer to~\cite{pa98} for discussions and results along this line.

In another approach, initiated in~\cite{pa76}, it was shown that to each
infinite class $A$, $B$, $C$ and $D$ of simple Lie algebras there
corresponds quantum statistics. Examples from classes $A$ and
$B$ of proper Lie superalgebras are also available. We have in
mind Wigner quantum systems (WQSs)~\cite{pa81}. Some such systems
possess quite unconventional physical features. As an example we
mention the $(n+1)$-particle WQS, based on the Lie superalgebra
$sl(1/3n)$ from class $A$~\cite{pa97}. This WQS exhibits a quark-like
structure~: the composite system occupies a small volume $V$
around the centre of mass and no particles can be extracted out
of $V$. Moreover the geometry within $V$ is noncommutative.
Another example is the $osp(3/2)$ WQS from class $B$~\cite{pa94}. It
leads to a picture where two spinless point particles, ``curling"
around each other, produce an orbital (internal angular) momentum
$1/2$, a result which cannot be obtained in 
canonical quantum mechanics.

The present paper is also in the frame of quantum statistics.
We study further the (microscopic) properties of $A$-statistics,
introduced in~\cite{pa76} (see also~\cite{pa77}), 
namely the statistics of Lie
algebras $A_n\equiv sl(n+1)$, $n=1,2,\ldots$.  Since Refs.~\cite{pa76} 
and~\cite{pa77} are not available as journal publications, we review the 
main issues of $A$-statistics in Sections~2 and~3, omitting most of
the proofs.

We begin (Section~2) by recalling how the Lie algebra $sl(n+1)$
can be described via generators $a_1^\pm,\ldots,a_n^\pm$ and triple
relations, see~(\ref{e2-5}). To the best of our knowledge such
generators were introduced for the first time by Jacobson~\cite{ja49}
in the more general context of Lie triple systems.
For this reason we refer to $a_1^\pm,\ldots,a_n^\pm$ as 
Jacobson generators (JGs). The latter provide an alternative to 
the Chevalley description of $sl(n+1)$.

The Fock modules of the Jacobson generators, extended also to
$gl(n+1)$-modules, are defined and classified in Section~3. It is
shown how they can be selected out of all irreducible
$gl(n+1)$-modules on the ground of natural physical requirements,
see Definition~\ref{def1}. 
All Fock modules $W_p$ are finite-dimensional
and are labelled by one positive integer $p\in \N$. Within $W_p$
each generator $a_i^+$ (resp.\ $a_i^-$) is interpreted as an
operator creating (resp.\ annihilating) a ``particle" in a state~$i$. 
The Pauli principle for $A$-statistics is also formulated
(Corollary~\ref{cor3}).

In Section~4 we argue that $A$-statistics can be interpreted as
a particular kind of exclusion statistics~\cite{ha91}.

Next, in Section~5, representation dependent creation and
annihilation operators $B(p)_i^\pm=a_i^\pm/ {\sqrt p}$
$(i=1,\ldots,n)$ in $W_p$ are defined. We prove that in an
appropriate topology $\displaystyle\lim_{p\to \infty} B(p)_i^\pm$
= $B_i^\pm$, where $B_1^\pm,\ldots,B_n^\pm$ are Bose creation and
annihilation operators. The operators $B(p)_1^\pm,\ldots,B(p)_n^\pm$
possess also other Bose-like properties. For these reasons
$B(p)_1^\pm,\ldots,B(p)_n^\pm$ are referred to as quasi-Bose
operators (of order $p$), the representations of $sl(n+1)$ and
$gl(n+1)$ in $W_p$ as quasiboson representations and the
statistics as quasi-Bose statistics.

The Jacobson CAOs $a_1^\pm,\ldots,a_n^\pm$ are ``bosonized" in 
Section~6. These operators are expressed via $n$ pairs of Bose CAOs
$B_1^\pm,\ldots,B_n^\pm$. The related realization of $gl(n+1)$ in
$W_p$ turns to be the known Holstein-Primakoff realization~\cite{ok75}.

In Section~7 we point out that the $p=1$ quasi-Bose operators
(coinciding in this case with the $p=1$ representation of the JGs)
can also be of more general interest. On the example of a two-leg
$S=1/2$ Heisenberg spin ladder we show that the Bose realization
of the Hamiltonian~\cite{go94, su98} together with the restrictions
selecting the physical subspace actually means that the Bose
operators related to each site  have to be replaced by quasi-Bose
operators of order $p=1$. This conclusion is of a more general
nature. It holds for any hard-core Bose model~\cite{fi89}, since the
$p=1$ quasibosons are hard-core bosons (Proposition~\ref{prop5}).

The final Section~8 is devoted to some conclusions and
discussions.

Throughout the paper we use the following abbreviations and
notation (some of them standard)~:
\begin{itemize}
\item[]
JGs -- Jacobson generators;
\item[]
CAOs -- creation and annihilation operators;
\item[]
UEA -- universal enveloping algebra;
\item[]
$\N$ -- all positive integers;
\item[]
$\Z_+$ -- all non-negative integers;
\item[]
$[a,b]=ab-ba,\qquad \{a,b\}=ab+ba$;
\item[]
$\oplus$, $\dot{\oplus}$ -- direct sum of linear
spaces and Lie algebras, respectively.
\end{itemize}

\section{Jacobson generators of $sl(n+1)$}
\setcounter{equation}{0}

The $sl(n+1)$-statistics, including $n=\infty$, was introduced 
in~\cite{pa76} (see also~\cite{pa77}) 
as an alternative way for quantization of
spinor fields in quantum field theory. Refs.~\cite{pa76} and~\cite{pa77}
are not available as journal publications. Therefore here and in
Section~3 we outline the main features of this statistics in
somewhat more details.

In order to define the Jacobson generators,
it is convenient to consider $A_n\equiv sl(n+1)$  as
a subalgebra of the Lie algebra $gl(n+1)$.  The universal
enveloping algebra $U[gl(n+1)]$ of the latter can be defined as
an associative algebra with unity of the Weyl generators
$\{e_{ij}|i,j=0,1,\ldots,n\}$ subject to the relations
\beq
[e_{ij},e_{kl}]=\delta_{jk}e_{il}-\delta_{il}e_{kj}.
\label{e2-1}
\eeq
Then $gl(n+1)$ is a subalgebra of $U[gl(n+1)]$, considered as a
Lie algebra, with generators $e_{ij}$, $i,j=0,1,\ldots,n$ and
commutation relations~(\ref{e2-1}).

The Cartan subalgebra $H'$ of $gl(n+1)$ has a basis
$h_{i}\equiv e_{ii}$, $i=0,1,\ldots,n$. Let $h^0,h^1, \ldots,h^n$
be the dual basis, $h^i(h_j)=\delta_{ij}$.
The root vectors of both $gl(n+1)$ and $sl(n+1)$ are
$e_{ij}$, $i\ne j=0,1,\ldots,n$.  The root of each $e_{ij}$ is
$h^{i} -h^{j}$. Then
\beq
sl(n+1)=\hbox{span}\{e_{ij}, e_{ii}-e_{jj}|i\ne j=0,1,\ldots,n\}. 
\label{e2-2}
\eeq
The Jacobson generators (JGs) of $sl(n+1)$ are part of
the Weyl generators, namely
\beq
a_i^+=e_{i0},\qquad a_i^-=e_{0i},\qquad i=1,\ldots,n. 
\label{e2-3}
\eeq
The correspondence with their roots reads
\beq
a_i^\pm \; \leftrightarrow \; \mp(h^0-h^i), \q i=1,\ldots,n,
\label{e2-4}
\eeq
and therefore the JGs $a_i^+$ ($a_i^-$) are negative
(positive) root vectors with respect to the natural ordering
$h^0,h^1,\ldots,h^n$. Since any other root is a sum of the roots of
$a_j^-$ and $a_i^+$, namely
\[
h^i-h^j=(h^0-h^j)-(h^0-h^i),\q  i\ne j=1,\ldots,n,
\]
the JGs~(\ref{e2-3}) generate $A_n$ in the sense of a Lie algebra.

{}From~(\ref{e2-1}) and~(\ref{e2-3}) one derives the triple relations
\bea
(a) && [[a_i^+,a_j^-],a_k^+]=\delta_{kj}a_i^+ +
\delta_{ij}a_k^+ , \nn\\
(b) && [[a_i^+,a_j^-],a_k^-]=-\delta_{ki}a_j^- - 
\delta_{ij}a_k^-, \label{e2-5}\\
(c) && [a_i^+,a_j^+]=[a_i^-,a_j^-]=0. \nn
\eea
On the contrary, setting $e_{ij}-\delta_{ij}e_{00}=[a_i^+,a_j^-]$,
one derives from~(\ref{e2-5}) the commutation relation between all
$sl(n+1)$ generators $e_{ij}$, $e_{ii}-e_{jj}$, $i\ne j=0,1,\ldots,n$.
The description of $A_n$ by means of
the generators~(\ref{e2-3}) and the 
relations~(\ref{e2-5}) was already given
in a paper by Jacobson~\cite{ja49}; therefore, the elements $a_i^\pm$
are referred to as Jacobson generators of $A_n$.

The above description of $sl(n+1)$ via generators and relations
is a particular case of describing Lie algebras via Lie
triple systems (LTSs), initiated by Jacobson~\cite{ja49} 
and further developed to the $\Z_2$-graded case by
Okubo~\cite{ok94}. Let us be more concrete.
By definition~\cite{ja49} a Lie triple system 
${\cal L}$ is a subspace of an associative
algebra $U$, so that ${\cal L}$ is closed under the ternary 
operation 
$\omega : {\cal L}\otimes {\cal L}\otimes {\cal L}\rightarrow 
{\cal L}$  defined as $\omega (a\otimes b \otimes c)=
[[a,b],c]$, $a,b,c\in {\cal L}$. 
The definiton of a Lie supertriple system 
(equivalent to the definition in~\cite{ok94}) is similar.
The difference is that
${\cal L}$ is a $\Z_2$-graded subspace of an associative superalgebra 
$U$ and the commutators in the definition of $\omega$ are 
replaced by supercommutators.

The JGs of $sl(n+1)$ are closely related to the above definition. 
More precisely, let  ${\cal L}_{sl}$ be the linear span of
the generators~(\ref{e2-3}) and $U_{sl}$ be the associative unital
algebra of the JGs subject to the relations~(\ref{e2-5}). Then
${\cal L}_{sl}$ is a subspace of $U_{sl}$. Moreover 
$\omega: {\cal L}_{sl}\otimes {\cal L}_{sl}\otimes 
{\cal L}_{sl}\rightarrow {\cal L}_{sl}$ as a consequence 
of~(\ref{e2-5}). Hence ${\cal L}_{sl}$ is a Lie triple system
with a basis consisting of the JGs~(\ref{e2-3}) and $U_{sl}$ is the
UEA of $sl(n+1)$. Similarly, the linear span ${\cal L}_{pf}$ 
of para-Fermi CAOs $F_1^\pm,F_2^\pm,\ldots,F_n^\pm$ together
with the associative algebra $U_{pf}$ of these operators 
(subject to the relations~(\ref{e1-4})) is another example of
a LTS.  Hence the para-Fermi operators $F_1^\pm,\ldots,F_n^\pm$
could be called JGs of $so(2n+1)$. In the same spirit
the para-Bose operators $B_1^\pm,\ldots,B_n^\pm$ are JGs 
of $osp(1/2n)$. 

{}From a purely algebraic point of view the
Jacobson generators provide an alternative
to the Chevalley description of $sl(n+1)$, $so(2n+1)$ and
$osp(1/2n)$. The JGs of  $so(2n+1)$ and  $osp(1/2n)$
however (contrary to the Chevalley generators)
have a direct physical significance. These operators
extend the canonical Fermi and Bose statistics  
to the more general parastatistics. 
Below we proceed to show that
the JGs of $sl(n+1)$ also introduce a new quantum
statistics, different from Bose and Fermi statistics and their
generalization -- parastatistics.
This statistics is intrinsically related to class $A$
of simple Lie algebras in the same way as the para-Fermi statistics is
related to class $B$ of simple Lie algebras.

Typically the ``commutation relations" between the creation and the
annihilation operators (or the related position and momentum
operators in case of finite degrees of freedom) are derived from
(more precisely, are required to be consistent with) {\it the
main quantization equation}
\beq
[H, a_i^\pm]=\pm \e_ia_i^\pm, \label{e2-6}
\eeq
where $H$ is the Hamiltonian and $i$ replaces all indices that
may appear (momentum, spin, charge, etc.). In quantum field
theory~(\ref{e2-6}) expresses the translation invariance of the field
(in infinitesimal form). In quantum mechanics the same equation
appears as a compatibility condition (in the sense of Wigner~\cite{wi50})
between the Heisenberg equations of motion and the
classical equations, if the system has a classical analogue (for
more details see~\cite{pa81,oh82}). There are certainly several other
conditions to be satisfied (Galilean or relativistic invariance,
causality, etc.; we refer to~\cite{pa97} for discussions in case of
noncanonical quantum mechanics). The possibility for choosing
different statistics essentially depends upon the way one
represents the Hamiltonian $H$. We are going to illustrate this
on the example of para-Fermi statistics.

Consider a nonrelativistic free field locked in a
finite volume. In the case of a Fermi field the Hamiltonian $\H$ is
written in a normal-product form
\beq
\H=\sum_i \e_i f_i^+ f_i^-, \label{e2-7}
\eeq
so that the energy of the vacuum is zero. Here $f_i^+$ ($f_i^-$)
are Fermi creation (annihilation) operators~: $ \{f_i^\xi,
f_j^\eta\}={1\over 4}(\xi-\eta)^2\delta_{ij}$, $\xi, \eta=\pm$
or $\pm 1$. Then~(\ref{e2-6}) holds,
\beq
[\H, f_i^\pm]=\pm \e_i f_i^\pm , \label{e2-8}
\eeq
and each $f_i^\xi$ can be interpreted as an operator creating
($\xi=+$) or annihilating ($\xi=-$) a particle, 
i.e.\ a fermion with energy $\e_i$. Eq.~(\ref{e2-8})
is not fulfilled however, if the Fermi operators in~(\ref{e2-7}) are
replaced by para-Fermi operators~(\ref{e1-4})~: 
for $H=\sum_i \e_i F_i^+ F_i^-$ the equation
\beq
[H, F_i^\pm] = \pm \e_i F_i^\pm  \label{e2-9}
\eeq
does not hold. Why? In order to answer this question using
proper Lie algebraic language assume that the sum in~(\ref{e2-7}) is
finite (finite number of Fermi oscillators),
\beq
\H=\sum_{i=1}^n \e_i f_i^+ f_i^-. \label{e2-10}
\eeq
This is only an intermediate step. The considerations below
remain valid for $n=\infty$. Recall now that any $n$ pairs of
Fermi CAOs generate a particular Fermi representation of the Lie
algebra $so(2n+1)\equiv B_n$, whereas the para-Fermi operators
$F_1^{\pm},\ldots,F_n^{\pm}$ are (representation independent)
generators of $so(2n+1)$~\cite{ka62, ry63}.  
Eq.~(\ref{e2-8}) is not preserved,
when passing to other representations of $B_n$, because $H$ is
not an element from $B_n$ and hence $[H, F_i^\pm]$ in the LHS 
of~(\ref{e2-9}) is not a representation independent commutator.  This
observation suggests also the answer~: 
one has to rewrite~(\ref{e2-10}) in
a representation independent form.  In order to achieve this,
represent~(\ref{e2-10}) in the following identical form~:
\beq
\H={1\over 2}\sum_{i=1}^n \e_i ([f_i^+,f_i^-] +
\{f_i^+,f_i^-\}).   \label{e2-11}
\eeq
Consider the Lie algebra generated from $f_1^\pm,\ldots,f_n^\pm$ and
$\{f_i^+,f_i^-\}$. Since $\{f_i^+,f_i^-\}=1$, we obtain a
representation of the Lie algebra $B_n\dot{\oplus} I$, where $I$
is the one-dimensional center. Now $\H  \in B_n\dot{\oplus} I $ and
therefore the commutation relations~(\ref{e2-8}) hold for any other
representation of $B_n\dot{\oplus} I$. In other words, if we
substitute $f_i^\pm \rightarrow F_i^\pm$ and
$\{f_i^+,f_i^-\} \rightarrow {\hat p}$ in~(\ref{e2-11}), i.e.~set
\beq
H={1\over 2}\sum_{i=1}^n \e_i ([F_i^+,F_i^-] + {\hat p}),
\label{e2-12}
\eeq
where ${\hat p}$ is a generator of the center $I$, then the
quantization condition~(\ref{e2-8}) will be fulfilled for any
representation of $B_n\dot{\oplus} I$ and in particular for the
para-Fermi operators~(\ref{e1-4})~: $[H, F_i^\pm]=\pm \e_i F_i^\pm$. The
requirement ${\hat p}|0\rangle=p|0\rangle$, 
$p\in \N $ (and $F_i^- F_j^+|0\rangle=\delta_{ij}p|0\rangle$, 
$F_i^- |0\rangle=0$), leads to a
representation with an order of the (para)statistics $p$~\cite{gr65}.
Then the energy of the vacuum is also zero.

We shall now apply a similar approach for the algebra $A_n$.
Let $E_{ij}$, $i,j=0,1,\ldots,n,$ be a square matrix 
of order $(n+1)$ with 1 on the
intersection of the $i^{th}$ row and the $j^{th}$ column and zeros
elsewhere, i.e.,
\beq
(E_{ij})_{kl}=\delta_{ik}\delta_{jl}, \; i,j=0,1,\ldots,n. 
\label{e2-13}
\eeq
The map $\pi : e_{ij} \rightarrow E_{ij}$, $i,j=0,1,\ldots,n,$
gives a representation of $gl(n+1)$ (usually referred to as the
defining representation). Its restriction to $A_n$ gives a
representation of $A_n$. The operators $ A_i^+=E_{i0}$,
$A_i^-=E_{0i}$, $i=1,2,\ldots,n$ satisfy the triple 
relations~(\ref{e2-5}). Set
\beq
{\hat H}=\sum_{i=1}^n \e_i A_i^+A_i^-.  \label{e2-14}
\eeq
Then
\beq
[{\hat H},A_i^\pm]=\pm \e_i A_i^\pm. \label{e2-15}
\eeq
Hence $A_i^\xi$ can be interpreted as an operator creating $(\xi=+)$ or
annihilating $(\xi=-)$ a particle (quasiparticle, excitation)
with energy $\e_i$ for any $i=1,\ldots,n$. The representation $\pi$
is an analog of the Fermi representation of para-Fermi
statistics.

The commutation relations~(\ref{e2-15}) do not hold for other
representations of $A_n$. In order to extend the class of
admissible representations we rewrite the Hamiltonian~(\ref{e2-14}), 
like in the Fermi case, in the following identical form
\beq
{\hat H}=\sum_{i=1}^n \e_i ([A_i^+,A_i^-]+E_{00}).  \label{e2-16}
\eeq
The Lie algebra generated from the operators $A_1^\pm,\ldots,A_n^\pm$
and $E_{00}$ is $gl(n+1)=A_n\dot{\oplus} I$ (in the
representation $\pi$). Since ${\hat H} \in gl(n+1)$ (in this
representation), (\ref{e2-15}) also holds for any other
representation of $gl(n+1)$. In other words the Hamiltonian
\beq
H=\sum_{i=1}^n \e_i ([a_i^+,a_i^-]+e_{00}) = \sum_{i=1}^n \e_i
([a_i^+,a_i^-]+h_{0})  \label{e2-17}
\eeq
satisfies~(\ref{e2-6}) for any other representation of
$gl(n+1)$.

One may argue that expression~(\ref{e2-17}) is not satisfactory,
because the Hamiltonian $H$ is not a function of the
Jacobson generators only. 
Below, in Corollary~\ref{cor1}, we show that within every
irreducible representation $H$ can be written as a function of
the JGs. Here we note that $[a_i^+,a_i^-]+e_{00} = h_i$ and
therefore the Hamiltonian~(\ref{e2-17}) can be represented manifestly as
an element from the Cartan subalgebra of $gl(n+1)$~:
\beq
H= \sum_{i=1}^n \e_i h_{i}.    \label{e2-18}
\eeq

\section{Fock representations of $sl(n+1)$}
\setcounter{equation}{0}

We proceed to outline those representations of the Jacobson
generators, which possess the main features of Fock space
representations in ordinary quantum theory.  In order to
distinguish between the abstract generators and their
representations, the JGs $a_1^\pm,\ldots,a_n^\pm$, considered as
operators in a certain $A_n$-module $W$, 
are called (Jacobson) creation and annihilation
operators of $A_n$ (abbreviated also as Jacobson CAOs of $A_n$,
$A_n$-CAOs, $A$-CAOs or simply CAOs).

\begin{defi}
Let $a_1^\xi,\ldots,a_n^\xi$ be Jacobson creation
$(\xi=+)$ and annihilation $(\xi=-)$ operators. The $A_n$-module
$W$ is said to be a Fock space of the algebra $A_n$ if it is a
Hilbert space, so that the following conditions hold~:
\begin{enumerate}
\item
Hermiticity condition ($A^*$ denotes the operator conjugate to $A$)
\beq
(a_i^+)^*=a_i^-, \q i=1,\ldots,n. \label{e3-1}
\eeq
\item
Existence of vacuum. There exists a vacuum vector
$|0\rangle \in W $ such that
\beq
a_i^-|0\rangle=0, \q i=1,\ldots,n. \label{e3-2}
\eeq
\item
Irreducibility. The representation space $W$ is spanned
on vectors
\beq
a_{i_1}^+a_{i_2}^+\cdots a_{i_m}^+|0\rangle, \q m\in \Z_+. 
\label{e3-3}
\eeq
\end{enumerate}
\label{def1}
\end{defi}

The Fock space of $A_n$ is also said to be an $A_n$-module of
Fock, a Fock module of the $A$-operators or simply a Fock space.

Assume that $W$ is a Fock space. Condition~(\ref{e3-1}) asserts
that any Fock representation is unitarizable with respect to this
star operation, considered as an antilinear antiinvolution on
$A_n$. It is known that all such representations are realized in
direct sums of finite-dimensional irreducible $A_n$-modules. 
Then~(\ref{e3-3}) yields that any Fock module is a finite-dimensional
irreducible $A_n$-module.

We list a few propositions, proofs of which can be
found in~\cite{pa76, pa77}.

\begin{prop}
The $A_n$-module $W$ is a Fock space if and only
if it is an irreducible finite-dimensional module with a highest
weight $\Lambda$ such that
\beq
a_i^-a_j^+x_\Lambda=0 \q i\ne j=1,\ldots,n. \label{e3-4}
\eeq
The vacuum $|0\rangle$ is unique (up to a multiplicative constant)
and can be identified with the highest weight vector $x_\Lambda$
in $W$~ $|0\rangle=x_\Lambda$.  
\label{prop1}
\end{prop}

Recall that the Hamiltonian $H$, see~(\ref{e2-18}), does not belong to
$A_n$.  It is an element from $gl(n+1$). In order to define $H$
as an operator in $W$, we extend each Fock module to an
irreducible $gl(n+1)$-module. To this end we define the action of
the $gl(n+1)$ central element (also $gl(n+1)$ Casimir operator)
$h_0+h_1+\ldots +h_n$ in $W$, setting
\beq
(h_0+h_1+\ldots +h_n)x=px \q \forall x\in W,  \label{e3-5}
\eeq
where $p$ can be any number.

The next proposition classifies the Fock spaces. Unless otherwise
stated, the roots and the weights are represented by their
coordinates in the basis $h^0,h^1,\ldots,h^n$, i.e.,
$
\lambda=\sum_{i=0}^n l_i h^i\equiv (l_0,l_1,\ldots,l_n).
$

\begin{prop}
The irreducible $gl(n+1)$-module $W_p$
is a Fock space, so that the energy of the vacuum is zero
($H|0\rangle=0$), if and only if its highest weight (namely the
weight of $|0\rangle$) is $\Lambda=ph^0\equiv(p,0,\ldots,0)$, i.e., if
\beq
h_0|0\rangle =p|0\rangle,\q h_i|0\rangle =0,\q i=1,\ldots,n, \label{e3-6}
\eeq
where $p$ is an arbitrary positive integer.
\label{prop2}
\end{prop}

{}From~(\ref{e2-3}) and~(\ref{e3-5}) $h_0+h_1+\cdots+h_n=p$,
$h_0-h_i=[a_i^-,a_i^+]$, $i=1,\ldots,n$, which yields
\beq
h_0={1\over{n+1}}\big(p+\sum_{i=1}^n [a_i^-,a_i^+]\big), \q
h_i={1\over{n+1}}\Big(p+n[a_i^+,a_i^-]- \sum_{k\ne i=1}^n
[a_k^+,a_k^-]\Big) \label{e3-7}
\eeq
The last result shows that within any Fock module the
Weyl generators $e_{ij}$ can be expressed as functions
of $a_1^\pm,\ldots,a_n^\pm$. In view of this we say that
$a_1^\pm,\ldots,a_n^\pm$ are Jacobson CAOs of both  $sl(n+1)$
and of $gl(n+1)$.

An immediate consequence of~(\ref{e2-17}) and~(\ref{e3-7}) is 
the following

\begin{coro}
Within every Fock module $W_p$ the
Hamiltonian~(\ref{e2-17}) can be expressed entirely via the Jacobson
creation and annihilation operators~:
\beq
H={1\over{n+1}}\sum_{i=1}^n \e_i\Big(p+n[a_i^+,a_i^-]- \sum_{k\ne
i=1}^n [a_k^+,a_k^-]\Big).  \label{e3-8}
\eeq
\label{cor1}
\end{coro}

{}From~(\ref{e3-4}), (\ref{e3-6}) and (\ref{e3-7}) one concludes~:

\begin{coro}
The Fock module $W_p$ with a highest
weight $\Lambda=(p,0,\ldots,0)$ is completely defined by the relations
\beq
a_i^-a_j^+|0\rangle=\delta_{ij}p|0\rangle, \q
a_k^-|0\rangle=0,\q p\in \N, \q
i,j,k=1,\ldots,n. \label{e3-9}
\eeq
\label{cor2}
\end{coro}

The above two conditions are the same as in the case of Green's
parastatistics of order $p$~\cite{gr53}. Therefore $p$ is referred to as
an order of $A_n$-statistics (or $A$-statistics). The
conclusion is that like in parastatistics the Fock spaces are
labelled by a positive integer $p\in\N$. The representations
corresponding to different orders of statistics have different
highest weights and are therefore inequivalent.

Taking into account the second relation $a_k^-|0\rangle=0$ 
in~(\ref{e3-9}),
one can also define the Fock module $W_p$ by means of the relations
\beq
[a_i^-a_j^+]|0\rangle=\delta_{ij}p|0\rangle, \q
a_k^-|0\rangle=0, \q p\in \N, \q i,j,k=1,\ldots,n. \label{e3-10}
\eeq
In view of this $A$-statistics and its Fock representations
can be formulated in a somewhat more mathematical terminology.
The latter is based on the observation that the linear span of
all generators $[a_i^-a_j^+]$, $a_i^-$, $i,j=1,\ldots,n,$ is a subalgebra
$\cal A$ of $gl(n+1)$ (which contains as subalgebra also
$gl(n)=\hbox{span} \{[a_i^-a_j^+]|i,j=1,\ldots,n \}$). 
Equations~(\ref{e3-10})
define one-dimensional representations of $\cal A$, spanned on the
vacuum $|0\rangle$. Therefore the Fock modules $W_p$ can be defined
as those irreducible finite-dimensional $gl(n+1)$-modules, which
are induced from trivial one-dimensional modules of $\cal A$ via
eqs.~(\ref{e3-10}). Then $p$ labels the different, inequivalent
one-dimensional modules of $\cal A$.

On the other hand one can define $A$-statistics by means of the
triple relations~(\ref{e2-5}). Then eqs.~(\ref{e3-9}) 
define completely the Fock
modules $W_p$. All calculations can be carried out without even
mentioning the underlying Lie algebraic structure of
$A$-statistics (which is usually the case for parastatistics).

Let $W_p$ be a Fock space with order of statistics $p$.
{}From~(\ref{e3-3}) and the fact that the creation operators
commute with each other one concludes that $W_p$ is a linear span
of vectors
$(a_1^+)^{l_1}(a_2^+)^{l_2}\cdots(a_n^+)^{l_n}|0\rangle$,
$l_1,\ldots,l_n\in
\Z_+$. The correspondence weight $\leftrightarrow$ weight vector
is one to one~:
\beq
(a_1^+)^{l_1}(a_2^+)^{l_2}\cdots(a_n^+)^{l_n}|0\rangle \q
\leftrightarrow \q
(p-\sum_{k=1}^n l_k,l_1,l_2,\ldots,l_n), \label{e3-11}
\eeq
i.e.\ all weight subspaces are one-dimensional.

\begin{prop}
Let $W_p$ be an $A_n$-module of Fock 
with order of statistics $p$. The vector
\beq
(a_1^+)^{l_1}(a_2^+)^{l_2}\cdots(a_n^+)^{l_n}|0\rangle  \label{e3-12}
\eeq
is not zero if and only if
\beq
l_1+l_2+\cdots+l_n\le p. \label{e3-13}
\eeq
\label{prop3}
\end{prop}

The proof is a consequence of the properties of the roots in any
finite-dimensional irreducible $A_n$-module $W$. If $\Lambda
=(L_0,L_1,\ldots,L_n)$ is the highest weight in $W$, then for any other
weight $\lambda =(l_0,l_1,\ldots,l_n)$ the following inequality holds~:
\beq
l_{i_0}+l_{i_1}+\cdots+l_{i_m} \le L_0+L_1+\cdots+L_m, \label{e3-14}
\eeq
where $i_0\ne i_1\ne \ldots,\ne i_m=0,1,\ldots,n$ and 
$m=0,1,\ldots,n$. Equation~(\ref{e3-14}) is an equality for $m=n$. 
If $W_p$ is a Fock space,
$L_0+L_1+\ldots+L_m=p$.

Proposition~\ref{prop3} can be proved also by a direct, but rather long
computation.  One verifies that the infinite-dimensional module
${\hat W_p}$ spanned on all vectors~(\ref{e3-12}) with $l_1,\ldots,l_n$ 
being arbitrary non-negative integers contains an invariant subspace
$V_p$ spanned on all vectors~(\ref{e3-12}) 
with $l_1+l_2+\ldots+l_n > p$.
Then $W_p$ is the factor module ${\hat W_p}/V_p$ and all 
vectors~(\ref{e3-12}), subject to~(\ref{e3-13}) are 
(representatives of) the basis
vectors in $W_p={\hat W_p}/V_p$.

We proceed to recall how one defines a metric in $W_p$, so that it
is a Hilbert space and the hermiticity condition~(\ref{e3-1}) holds.
Consider the vectors
\beq
(a_1^+)^{l_1}(a_2^+)^{l_2}\cdots(a_n^+)^{l_n}|0\rangle, \q
l_1+l_2+\cdots+l_n\le p  \label{e3-15}
\eeq
from $W_p$.  All such vectors have different weights.
Consequently they are linearly independent and can be considered
as a basis in $W_p$. Define a Hermitian form $(~,~)$ on $W_p$ in
the usual way (for quantum theory), postulating (in addition to
$a_i^-|0\rangle=0$, see~(\ref{e3-2}))~:
\bea
(a) &\ & (|0\rangle,|0\rangle)\equiv \langle 0|0\rangle=1, \nn\\
(b) && \langle 0|a_i^+ =0, \q i=1,\ldots,n, \label{e3-16}\\
(c) && \left((a_1^+)^{m_1}(a_2^+)^{m_2}\cdots(a_n^+)^{m_n}|0\rangle,
     (a_1^+)^{l_1}(a_2^+)^{l_2}\cdots(a_n^+)^{l_n}|0\rangle\right)=\nn\\
&&   \langle 0| (a_n^-)^{m_n}\cdots(a_2^-)^{m_2}(a_1^-)^{m_1}
     (a_1^+)^{l_1}(a_2^+)^{l_2}\cdots(a_n^+)^{l_n}|0\rangle.\nn
\eea
With respect to this form the vectors~(\ref{e3-15}) are orthogonal.
Moreover,
\beq
\Big((a_1^+)^{l_1}(a_2^+)^{l_2}\cdots(a_n^+)^{l_n}|0\rangle,~
(a_1^+)^{l_1}(a_2^+)^{l_2}\cdots(a_n^+)^{l_n}|0\rangle \Big)= 
{p!\over
(p-\sum_{j=1}^n l_j )!}\prod_{i=1}^n l_i!>0.  \label{e3-17}
\eeq
Therefore all vectors
\beq
|p;l_1,\ldots,l_n\rangle=\sqrt{(p-\sum_{j=1}^n l_j )!\over p!}
{(a_1^+)^{l_1}\ldots(a_n^+)^{l_n}\over{\sqrt{l_1!l_2!\ldots
l_n!}}}|0\rangle, \q l_1+l_2+\cdots+l_n\le p 
\label{e3-18}
\eeq
constitute an orthonormal basis in $W_p$, i.e.\ $(~,~)$ is a
scalar product. Then by construction the hermiticity 
condition~(\ref{e3-1}) holds too.

The transformation of the basis~(\ref{e3-18}) under the action of the
Jacobson CAOs reads~:
\bea
a_i^+|p;l_1,\ldots ,l_i,\ldots,l_n\rangle&=&
  \sqrt{(l_i+1)(p-\sum_{j=1}^n l_j  )}~
  |p;l_1,\ldots,l_{i-1},l_i+1,l_{i+1}\ldots,l_n\rangle, 
\label{e3-19a}\\
a_i^-|p;l_1,\ldots,l_i,\ldots,l_n\rangle&=&
  \sqrt{l_i(p-\sum_{j=1}^n l_j +1  )}~
  |p;l_1,\ldots,l_{i-1},l_i-1,l_{i+1}\ldots,l_n\rangle. 
\label{e3-19b}
\eea
Moreover,
\bea
&& h_0 |p;l_1,l_2,\ldots,l_n\rangle=(p-\sum_{i=1}^n
l_i)|p;l_1,l_2,\ldots,l_n\rangle, \label{e3-20a} \\
&& h_i |p;l_1,l_2,\ldots,l_n\rangle=l_i|p;l_1,l_2,\ldots,l_n\rangle, 
\q i=1,\ldots,n. \label{e3-20b}
\eea

Let us consider in some more detail
the $p=1$ representation.
Denote by $b_i^\pm$ the Jacobson CAOs $a_i^\pm$ in this representation.
In this particular case the
representation space $W_1$ is $(n+1)$-dimensional with a basis
\beq
|1;l_1,\ldots,l_n\rangle, \q l_1+\cdots+l_n\le 1, \label{e3-21}
\eeq
i.e.\ at most one of the labels $l_1,\ldots,l_n$ in
$|1;l_1,\ldots,l_n\rangle$ is equal to 1 and all other are zeros.  
Then~(\ref{e3-19a})-(\ref{e3-19b}) reduces to
\beq
\begin{array}{l}
b_i^+|1;l_1,\ldots,l_{i-1},l_i,l_{i+1},\ldots,l_n\rangle
  =(1-l_i)|1;l_1,\ldots,l_{i-1},l_i+1,l_{i+1},\ldots,l_n\rangle, \\[2mm]
b_i^-|1;l_1,\ldots,l_{i-1},l_i,l_{i+1},\ldots,l_n\rangle=
  l_i|1;l_1,\ldots,l_{i-1},l_i-1,l_{i+1},\ldots,l_n\rangle. 
\end{array}
\label{e3-22}
\eeq
The matrix elements of $b_i^+$ and $b_i^-$, in the basis ordered
as $|1;0,0,0,\ldots,0\rangle$, $|1;1,0,0,\ldots,0\rangle$, 
$|1;0,1,0,\ldots,0\rangle$,
$|1;0,0,1,\ldots,0\rangle$, $\ldots$, $|1;0,0,0,\ldots,1\rangle$ 
are the same as
those of the Weyl generators $E_{i 0}$ and $E_{0 i }$ in the
defining $(n+1)$-dimensional matrix representation, see~(\ref{e2-13}).
Hence the $p=1$ representation is the same as the defining
representation and one can think of the operators $b_{i}^\pm$ as
of matrices,
\beq
E_{i 0}=b_{i}^+, \q E_{0 i}=b_{i}^-, \q i=1,\ldots,n.  \label{e3-23}
\eeq
{}From here and~(\ref{e3-7}) (with $p=1$) one can express also the
rest of the Weyl generators~(\ref{e2-13}) via $p=1$ Jacobson
creation and annihilation operators~:
\beq
E_{00}={1\over{n+1}}(1-\sum_{i=1}^n [b_{i}^+,b_{i}^-]), \q
E_{ij}=[b_{i}^+,b_{j}^-]+ {\delta_{ij}\over{n+1}} (1-\sum_{k=1}^n
[b_{k}^+,b_{k}^-]), \q
i,j=1,\ldots,n. \label{e3-24}
\eeq

\section{The Pauli principle for $A$-statistics}
\setcounter{equation}{0}

The results obtained so far justify the terminology used. 
Equations~(\ref{e2-18}) and~(\ref{e3-6}) yield
\beq
H|p;l_1,\ldots,l_i,\ldots,l_n\rangle =\sum_{i=1}^n
l_i\e_i|p;l_1,\ldots,l_i,\ldots,l_n\rangle.  \label{e4-1}
\eeq
Therefore the state $|p;l_1,\ldots,l_i,\ldots,l_n\rangle$ 
can be interpreted
as a many-particle state with $l_1$ particles on the first
orbital, $l_2$ particles on the second orbital, etc. For reasons 
that will become clear soon, we refer to these particles as 
quasibosons (of order $p$).
The operator
$h_i$, $i=1,\ldots,n,$ see~(\ref{e3-20b}), 
is the number operator for the
quasibosons on the $i^{th}$ orbital, whereas  ${\hat N}=h_1+\cdots+h_n$
counts all quasibosons, accommodated in the state
$|p;l_1,\ldots,l_i,\ldots,l_n\rangle$.

Since, see~(\ref{e3-19a}),
\beq
a_i^+|p;l_1,\ldots ,l_i,\ldots,l_n\rangle \sim |p;l_1,\ldots,
l_{i-1},l_i+1,l_{i+1},\ldots,l_n\rangle, \ 
\hbox{ if }\ \sum_{i=1}^n l_i<p,   \label{e4-2}
\eeq
the operator $a_i^+$ creates a quasiboson on the $i^{th}$ orbital,
a particle with energy $\e_i$, if the state contains
less than $p$ quasibosons. On the other hand,
$a_i^+|p;l_1,\ldots, l_{i-1},l_i,l_{i+1},\ldots,l_n\rangle=0$,
if $\sum_{i=1}^n l_i=p$, i.e.\ no more than $p$ quasibosons can be
accommodated.  Similarly,  if $l_i>0$, $a_i^-$ ``kills" a quasiboson
with energy $\e_i$. Therefore, reformulating Proposition~\ref{prop3}, 
one obtains~:

\begin{coro}[Pauli principle for $A$-statistics]
Let $W_p$ be a Fock space of $A_n$, corresponding to an order of
statistics $p$.  Within $W_p$ all states containing no more than
$p$ quasibosons, namely all states
\beq
|p;l_1,\ldots,l_i,\ldots,l_n\rangle\ \hbox{ with }\ 
0\le\sum_{i=1}^n l_i \le p,
\label{e4-3}
\eeq
are allowed. There are no states accommodating more than $p$
particles.
\label{cor3}
\end{coro}

Let us consider, as an example, $A$-statistics of order $p=4$
with $n=6$ orbitals (for instance with 6 different energy levels).
{}From~(\ref{e4-3}), it follows that there is no restriction
on the number of quasibosons to be accommodated on a certain 
orbital as long as the total number of particles in 
any configuration does not exceed $p$.
Hence, the following three states or configurations are allowed
(the orbitals,
for instance the energy levels, are represented by lines, 
and the quasibosons by dots)~:
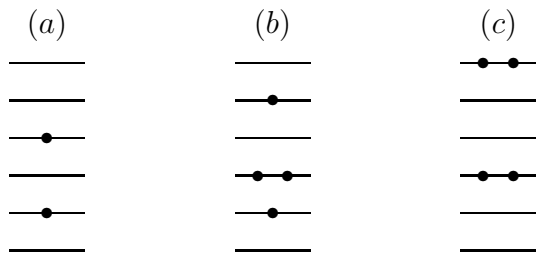
\[
\vbox{
\unitlength=1.00mm
\special{em:linewidth 0.4pt}
\linethickness{0.4pt}
\begin{picture}(80.00,150.00)
\put(10.00,145.00){\line(1,0){10.00}}
\put(40.00,145.00){\line(1,0){10.00}}
\put(70.00,145.00){\line(1,0){10.00}}
\put(10.00,140.00){\line(1,0){10.00}}
\put(40.00,140.00){\line(1,0){10.00}}
\put(70.00,140.00){\line(1,0){10.00}}
\put(10.00,135.00){\line(1,0){10.00}}
\put(40.00,135.00){\line(1,0){10.00}}
\put(70.00,135.00){\line(1,0){10.00}}
\put(10.00,130.00){\line(1,0){10.00}}
\put(40.00,130.00){\line(1,0){10.00}}
\put(70.00,130.00){\line(1,0){10.00}}
\put(10.00,125.00){\line(1,0){10.00}}
\put(40.00,125.00){\line(1,0){10.00}}
\put(70.00,125.00){\line(1,0){10.00}}
\put(10.00,120.00){\line(1,0){10.00}}
\put(40.00,120.00){\line(1,0){10.00}}
\put(70.00,120.00){\line(1,0){10.00}}
\put(15.00,150.00){\makebox(0,0)[cc]{$(a)$}}
\put(45.00,150.00){\makebox(0,0)[cc]{$(b)$}}
\put(75.00,150.00){\makebox(0,0)[cc]{$(c)$}}
\put(15,125){\circle*{1.5}}
\put(15,135){\circle*{1.5}}
\put(45,125){\circle*{1.5}}
\put(43,130){\circle*{1.5}}
\put(47,130){\circle*{1.5}}
\put(45,140){\circle*{1.5}}
\put(73,130){\circle*{1.5}}
\put(77,130){\circle*{1.5}}
\put(73,145){\circle*{1.5}}
\put(77,145){\circle*{1.5}}
\end{picture}
\vskip -122mm
}
\]
Note that the last two configurations $(b)$ and $(c)$ are already 
``saturated'' in
the sense that no more quasibosons can be added, since the total
number of particles is already equal to $p=4$.
The following two configurations correspond to
forbidden states~:
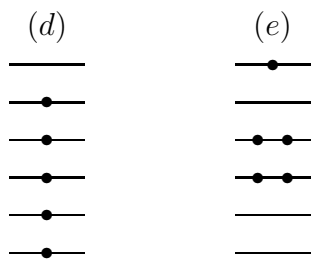
\[
\vbox{
\unitlength=1.00mm
\special{em:linewidth 0.4pt}
\linethickness{0.4pt}
\begin{picture}(80.00,150.00)
\put(10.00,145.00){\line(1,0){10.00}}
\put(40.00,145.00){\line(1,0){10.00}}
\put(10.00,140.00){\line(1,0){10.00}}
\put(40.00,140.00){\line(1,0){10.00}}
\put(10.00,135.00){\line(1,0){10.00}}
\put(40.00,135.00){\line(1,0){10.00}}
\put(10.00,130.00){\line(1,0){10.00}}
\put(40.00,130.00){\line(1,0){10.00}}
\put(10.00,125.00){\line(1,0){10.00}}
\put(40.00,125.00){\line(1,0){10.00}}
\put(10.00,120.00){\line(1,0){10.00}}
\put(40.00,120.00){\line(1,0){10.00}}
\put(15.00,150.00){\makebox(0,0)[cc]{$(d)$}}
\put(45.00,150.00){\makebox(0,0)[cc]{$(e)$}}
\put(15,120){\circle*{1.5}}
\put(15,125){\circle*{1.5}}
\put(15,130){\circle*{1.5}}
\put(15,135){\circle*{1.5}}
\put(15,140){\circle*{1.5}}
\put(43,130){\circle*{1.5}}
\put(47,130){\circle*{1.5}}
\put(43,135){\circle*{1.5}}
\put(47,135){\circle*{1.5}}
\put(45,145){\circle*{1.5}}
\end{picture}
\vskip -122mm
}
\]
None of the states $(d)$ and $(e)$ is allowed since the total number of
particles in the configuration exceeds $p=4$.

This example clearly illustrates the accommodation properties of
$A$-statistics of order $p$. Because of this ``exclusion principle'',
$A$-statistics can be interpreted as 
a special case of exclusion statistics
in the sense of Wu~\cite{wu94}. We recall that
\beq
d(N)=n-g\cdot(N-1). 
\label{e4-4}
\eeq
This should be interpreted as follows~: 
let $n$ be the total number of orbitals
that are available for the first particle, 
and suppose $N-1$ particles are
already accommodated in the configuration, then $d(N)$ expresses the
dimension of the single-particle space for the $N^{\rm th}$ particle (or
the number of orbitals where the $N^{\rm th}$ particle can be ``loaded'').
Bose statistics has $g=0$, and Fermi statistics has $g=1$.

If one accepts the natural assumption that~(\ref{e4-4}) 
should hold for all {\it admissible}
values of $N$, i.e.\ one does 
not require~(\ref{e4-4}) to be applicable for values of $N$
which the system cannot accommodate, then
$A$-statistics is a particular case of exclusion statistics, 
also with $g=0$~: 
\beq
d(N)=n, \q \forall\, N \in \{1,2,\ldots,p\}. \label{e4-5}
\eeq
$A$-statistics is similar to Bose statistics in the sense that
there is no restriction on the number of particles on an orbital. 
The main difference comes from the fact that the total configuration 
should contain
no more than $p$ particles. This is one of the reasons
(see also next Section~5) to refer to
$A$-statistics as quasi-Bose statistics and to the particles 
as quasibosons.

If one drops the condition for
$N$ to be an admissible value, one
cannot satisfy equation~(\ref{e4-4}). Indeed,
(\ref{e4-4}) with $g=0$, does not hold for $N=p+1$, since 
$d(p+1)=0$~\cite{me98, me99}.

\section{Quasi-Bose creation and annihilation operators}
\setcounter{equation}{0}

In the present section we show first approximately and then in a
strict sense that $A$-statistics can be viewed as a
good finite-dimensional approximation to Bose statistics for
large values of order of statistics $p$. The terminology {\it
finite-dimensional approximation} comes to remind that the Fock
spaces $W_p$ of $A$-statistics are finite-dimensional linear
spaces, whereas any Bose Fock space is infinite-dimensional.

Introduce new, representation dependent, creation and 
annihilation operators
\beq
B(p)_i^\pm = {a_i^\pm\over \sqrt{p}}, \q i=1,\ldots,n, \q p\in \N,
\label{e5-1}
\eeq
in $W_p$. The transformations following 
from~(\ref{e3-19a})-(\ref{e3-19b}) read~:
\bea
&& B(p)_i^+|p;l_1,\ldots,l_i,\ldots,l_n\rangle=
  \sqrt{(l_i+1)(1-{{\sum_{k=1}^n l_k}\over p})}~
|p;l_1,\ldots,l_i+1,\ldots,l_n\rangle, \label{e5-2a} \\
&& B(p)_i^-|p;l_1,\ldots,l_i,\ldots,l_n\rangle=
  \sqrt{l_i(1+{{1-\sum_{k=1}^n l_k}\over p})}~
|p;l_1,\ldots,l_i-1,\ldots,l_n\rangle. \label{e5-2b}
\eea
Consider the above equations for values of the order of
statistics $p$, which are much greater than the number of
accommodated quasibosons, namely $l_1+l_2+\cdots+l_n \ll p.$ In this
approximation one obtains~:
\beq
\begin{array}{l}
B(p)_i^-|p;l_1,\ldots, l_{i-1},l_i,l_{i+1},\ldots,l_n\rangle \simeq
 \sqrt{l_i}\; |p;l_1,\ldots,l_{i-1},l_i-1,l_{i+1},\ldots,l_n\rangle,\\[2mm]
B(p)_i^+|p;l_1,\ldots, l_{i-1},l_i,l_{i+1},\ldots,l_n\rangle \simeq
 \sqrt{l_i+1}\;|p;l_1,\ldots,l_{i-1},l_i+1,l_{i+1}\ldots,l_n\rangle, 
\end{array}
\label{e5-3}
\eeq
which yields (an approximation to) the Bose commutation relations~:
\bea
&& [B(p)_i^+,B(p)_j^+]=[B(p)_i^-,B(p)_j^-]=0,\q 
\hbox{(exact commutators)}, \label{e54-a} \\
&& [B(p)_i^-,B(p)_j^+]\simeq \delta_{ij}, 
\q \hbox{ if } l_1+l_2+\cdots+l_n \ll p.   \label{e5-4b}
\eea
Since for $l_1+l_2+\cdots+l_n\equiv\sum_k l_k \ll p$
\[
{{(p-\sum_k l_k)!}\over{p!}}~p^{\sum_k l_k}=
{p\over{p-\sum_k l_k+1}}~{p\over{p-\sum_k l_k+2}}\ldots{p\over
p}\simeq 1,
\]
in a first approximation~(\ref{e3-18}) reduces also to the well known
expressions for the orthonormed basis in a Fock space of $n$
pairs of Bose creation and annihilation operators~:
\beq
|p;l_1,\ldots,l_n\rangle=
{(B(p)_1^+)^{l_1}\cdots(B(p)_n^+)^{l_n}\over{\sqrt{l_1!l_2!\cdots
l_n!}}}|0\rangle. 
\label{e5-5}
\eeq
The conclusion is that the representations of $B(p)_i^\pm$ in
(finite-dimensional)
state spaces $W_p$ with large values of $p$, restricted to states
with a small amount $l_1+l_2+\cdots+l_n \ll p$ of accommodated
quasibosons, provide a good approximation to Bose creation and
annihilation operators~\cite{pa76, pa77}.
This is another reason to 
refer to the operators $B(p)_i^\pm$ as 
{\it quasi-Bose creation and annihilation operators (of order $p$)}.

In the remaining part of this section we will prove that in the
limit $p \rightarrow \infty$ the quasi-Bose operators reduce to
Bose creation and annihilation operators. To this end we proceed
to introduce first an appropriate topology.

Let $W$ be a Hilbert space with an orthonormed basis
\beq
|l_1,\ldots,l_i,\ldots,l_n\rangle\equiv |L\rangle, \q \forall\, 
l_1,\ldots,l_n \in \Z_+.
\label{e5-6}
\eeq
Whenever possible we write $|L\rangle$ as an abbreviation 
for $|l_1,\ldots,l_i,\ldots,l_n\rangle$ and denote
by $|L\rangle_{\pm i}$ a vector obtained from $|L\rangle$
by replacing $l_i$ with $l_i\pm 1$, namely
\beq
|L\rangle_{\pm i}=|l_1,\ldots,l_{i-1},l_i\pm 1,l_{i+1},
\ldots,l_n\rangle.
\label{e5-7}
\eeq
The space $W$ consists of all vectors
\beq
\Phi=\sum_{l_1=0}^\infty \cdots \sum_{l_n=0}^\infty
c(l_1,\ldots,l_n)|l_1,\ldots,l_n\rangle \equiv \sum_{L}c(L)|L\rangle, 
\label{e5-8}
\eeq
where $c(l_1,\ldots,l_n)\equiv c(L)$ are complex numbers such that
\beq
\sum_{l_1=0}^\infty \cdots \sum_{l_n=0}^\infty |c(l_1,\ldots,l_n)|^2
\equiv \sum_{l_1,\ldots,l_n=0}^\infty |c(l_1,\ldots,l_n)|^2 \equiv
\sum_{L} |c(L)|^2 < \infty ,
\label{e5-9}
\eeq
and this is in fact the square of the Hilbert space norm $(|\Phi|_0)^2$
of $\Phi$.

Embed the $sl(n+1)$-module $W_p$ in $W$ via an identification
of the basis vectors
\beq
|p;l_1,\ldots,l_i,\ldots,l_n\rangle\equiv 
|l_1,\ldots,l_i,\ldots,l_n\rangle\equiv
|L\rangle \q \forall~ l_1+\ldots+l_n\le p. 
\label{e5-10}
\eeq
In order to turn the entire space $W$ into an
$sl(n+1)$-module, so that the restriction on $W_p\subset W$
coincides with~(\ref{e5-2a})-(\ref{e5-2b}), we set~:
\bea
&& B(p)_i^+\Phi=\sum_{l_1+\cdots+l_n \le p}c(L)
\sqrt{(l_i+1)(1-{{\sum_{k=1}^n l_k}\over p})}~|L\rangle_i,
\label{e5-11a} \\
&& B(p)_i^-\Phi=\sum_{l_1+\cdots+l_n \le p}c(L)
\sqrt{l_i(1+{{1-\sum_{k=1}^n l_k}\over
p})}~|L\rangle_{-i},
\label{e5-11b}
\eea
where $\Phi$ is any vector~(\ref{e5-8}) from $W$ and 
$\sum_{l_1+\cdots+l_n
\le p}$ is a sum over all possible $l_1,\ldots,l_n \in \Z_+$ such that
$l_1+\cdots+l_n \le p$. Note that the sums 
in~(\ref{e5-11a})-(\ref{e5-11b}) are finite.

The transformation of the basis, following 
from~(\ref{e5-11a})-(\ref{e5-11b}), reads~:
\bea
&& B(p)_i^+|L\rangle=
  \sqrt{(l_i+1)(1-{{\sum_{k=1}^n l_k}\over p})}~ |L\rangle_{i},
\q \forall~L\hbox{ such that }{{\sum_{k=1}^n l_k}}\le p, 
\label{e5-12a}\\
&& B(p)_i^-|L\rangle=
  \sqrt{l_i(1+{{1-\sum_{k=1}^n l_k}\over p})}~|L\rangle_{-i},
\q \forall~L\hbox{ such that }{{\sum_{k=1}^n l_k}}\le p,
\label{e5-12b}\\
&& B(p)_i^\pm|L\rangle=0,\q \forall~L\hbox{ such that }
{{\sum_{k=1}^n l_k}}> p.
\label{e5-12c}
\eea
The relations~(\ref{e5-12a})-(\ref{e5-12b}) are the same 
as~(\ref{e5-2a})-(\ref{e5-2b}) (via the
identification~(\ref{e5-10})).

Since the quasi-Bose operators $B(p)_i^\pm$ take values in a
finite-dimensional subspace of $W$, 
see~(\ref{e5-11a})-(\ref{e5-11b}), they are bounded
and hence continuous linear operators in $W$.
In view of this, see~(\ref{e5-8}), $ B(p)_i^\pm \Phi=B(p)_i^\pm
\sum_{L}c(L)|L\rangle=\sum_{L}c(L)B(p)_i^\pm|L\rangle $ and 
therefore~(\ref{e5-11a})-(\ref{e5-11b}) are a consequence 
of~(\ref{e5-12a})-(\ref{e5-12c}).

Next we proceed to define $n$ pairs of Bose operators
$B_i^\pm$, $i=1,\ldots,n$, in $W$. It is known that such operators
cannot be realized as bounded operators in $W$ (so that the
corresponding position and momentum operators  are selfadjoint
operators in $W$;
see, for instance, \cite{re72} or~\cite{em72}). 
Therefore care has to be taken about the common domain of
definition $\O$ of the Bose operators. Following~\cite{bo75} we set $\O$
to be a dense subspace of $W$
(with respect to the Hilbert space topology), 
consisting of all vectors~(\ref{e5-8})
for which the series
\beq
(|\Phi|_r)^2=\sum_{l_1,\ldots,l_n=0}^\infty (1+\sum_{k=1}^n
l_k)^r|c(l_1,\ldots,l_n)|^2 
\label{e5-13}
\eeq
is convergent for any $r=0,1,2,\ldots$. Then the relations
\beq
B_i^-|L\rangle =  \sqrt{l_i}\; |L\rangle_{-i}, \qquad
B_i^+|L\rangle =  \sqrt{l_i+1}\;|L\rangle_{i},
\label{e5-14}
\eeq
define a representation of $n$ pairs of bosons
$B_1^\pm,\ldots,B_n^\pm$, namely of operators, 
which satisfy the relations
\beq
[B_i^-,B_j^+]= \delta_{ij},\q
[B_i^+,B_j^+]=[B_i^-,B_j^-]=0, \q
i,j=1,\ldots,n, 
\label{e5-15}
\eeq
in $\Omega$ (with $\Omega$ being a common domain of definition for all
them). In terms of these operators
\beq
(|\Phi|_r)^2=(\Phi,(1+\sum_{k=1}^n B_k^+B_k^-)^r\Phi). 
\label{e5-16}
\eeq
The norms $|\Phi|_r$, $r=0,1,2,\ldots,$ turn
$\Omega$ into a countably normed topological space (which can be
viewed also as a metric space~\cite{ge68}).
All balls
\beq
B(\Phi_0;r,\epsilon)=\{\Phi\in \O~|~|\Phi-\Phi_0|_r < \epsilon \},
\q\forall~\Phi_0\in \O, \q\forall~r\in \Z_+, \q\forall~\epsilon >0,
\label{e5-17}
\eeq
constitute a basis of open sets in the countably normed
topological space $\O$, whereas the balls~(\ref{e5-17}) with a fixed $r$
yield a basis in $\O$, viewed as a $|\,\cdot\,|_r$-normed topological
space. Clearly any $|\,\cdot\,|_r$-normed topology ({\it 
$r$-normed topology})
is weaker than the countably normed topology ({\it $cn$-topology}).

{}From now on we restrict the domain of definition of all
quasi-Bose operators~(\ref{e5-1}) to be $\O$.  The fact that each
quasi-Bose operator maps $\O$ into a finite-dimensional subspace
of $\O$, see~(\ref{e5-11a})-(\ref{e5-11b}), 
indicates that each such operator is a
bounded and hence a continuous linear operator with respect to the
$r$-normed topology for any $r\in\Z_+$. A similar property however
does not hold for the Bose creation and annihilation 
operators~(\ref{e5-14}). 
These operators are not continuous with respect to any of
the $r$-normed topologies in $\O$. Therefore, if
$\sum_{i=1}^\infty \Phi_i=\Phi$ converges in the sense of a certain
$r$-normed topology, for instance in the Hilbert space topology
($r=0$), one cannot in general use relations like
\beq
B_i^\pm \sum_{i=1}^\infty \Phi_i= \sum_{i=1}^\infty B_i^\pm\Phi_i.
\label{e5-18}
\eeq
One of the advantages of the $cn$-topology is that it avoids the
above difficulties.  Here are some of the properties of this
topology, which will be relevant for the rest of the 
exposition~\cite{bo75}~:
\begin{itemize}
\item
$\Omega$ is stable under the action of any polynomial of Bose
operators, 
\beq
P(B_1^\pm,\ldots,B_n^\pm)\Omega~\subset~\Omega; 
\label{e5-19a}
\eeq
\item
Any polynomial of Bose CAOs is a continuous linear
operator in $\Omega$ with respect to the
$cn-$topology;
\vskip -11mm \beq \label{e5-19b} \eeq
\item
The scalar product in $\Omega$ is continuous with
respect to the convergence defined by the $cn$-topology.
\vskip -11mm \beq \label{e5-19c} \eeq
\end{itemize}

As a consequence, (\ref{e5-18}) holds for any series
$\sum_{i=1}^\infty \Phi_i$ which converges in the $cn$-topology;
moreover~(\ref{e5-19c}) yields $(\sum_{i=1}^\infty
\Phi_i,\Psi)=\sum_{i=1}^\infty (\Phi_i,\Psi)$. 
The relevance of
the $cn$-topology however goes far beyond the above
considerations. This topology, called nuclear topology, is of
prime importance in the theory of generalized functions~\cite{ge68,ge64},
and their applications in quantum theory (see, for instance~\cite{bo75}).

Let $\P$ be the set of all linear operators in $\O$ defined
everywhere in $\Omega$, which are continuous in the
$cn$-topology. With respect to the usual operations between
operators $\P$ is an associative algebra~\cite{ge68}.
According to~(\ref{e5-19b}) the Bose operators belong to $\P$.
The quasi-Bose operators~(\ref{e5-1}) (with domain of definition
restricted to $\Omega$) also belong to $\P$.  Indeed $B(p)_i^\pm$
are bounded and hence continuous operators in $\O$ with respect
to any $r$-normed topology. Let $B(\Phi_0;r,\epsilon)$ be an arbitrary
open ball in the $cn-$topology, see~(\ref{e5-17}). 
$B(\Phi_0;r,\epsilon)$ is an
open ball also in the $r$-normed topology. Therefore the inverse
image $O=[B(p)_i^\pm]^{-1}B(\Phi_0;r,\epsilon)$ 
of $B(\Phi_0;r,\epsilon)$ is
an open set in the $r$-normed topology. Since the latter is weaker
than the $cn$-topology, $O$ is an open set also in the
$cn$-topology.  Thus, the inverse image
$O=[B(p)_i^\pm]^{-1}B(\Phi_0;r,\epsilon)$ of any open ball 
(i.e.\ of any
open set from the basis) in the $cn$-topology is an open set with
respect to the same topology.  Therefore $B(p)_i^\pm$ is a
continuous operator in the $cn$-topology.

Introduce a topology on $\P$ in a way similar to the strong
topology in the algebra ${\cal B(H)}$ of all bounded linear
operators on a Hilbert space ${\cal H}$~\cite{na72}.  Let
$\Phi_1,\ldots,\Phi_s$ be $s$ different elements from $\Omega$
and $\epsilon$ be a positive number. A strong neighborhood
$U(A_0;\Phi_1,\ldots,\Phi_s;\epsilon)$ of the operator $A_0\in \P$ is
(defined as) the set of all operators $A\in \P$, which satisfy
the inequalities
\beq
|(A-A_0)\Phi_k|_0<\epsilon,\q \forall\,k=1,\ldots,s.  
\label{e5-20}
\eeq

\begin{defi}
A {\em strong topology} on $\P$ is the
topology with a basis of open sets consisting of all possible
strong neighborhoods $U(A_0;\Phi_1,\ldots,\Phi_s;\epsilon)$ (namely the
collection of strong neighborhoods, corresponding to any $A_0\in
\P$, to any $\epsilon>0$, to any $s\in \N$ and to any sequence
$\Phi_1,\ldots,\Phi_s$ of different elements from $\O$).
\label{def2}
\end{defi}

\begin{prop}
In the strong topology
\beq
\displaystyle\lim_{p\to \infty} B(p)_i^\pm =B_i^\pm,\q i=1,\ldots,n.
\label{e5-21}
\eeq
\label{prop4}
\end{prop}

\noindent {\it Proof.} 
In order to prove that~(\ref{e5-21}) holds it is sufficient
to show that every strong neighborhood
$U(B_i^\pm;\Phi_1,\ldots,\Phi_s;\epsilon)$ of $B_i^\pm$ contains all
elements of the sequence $B(1)_i^\pm,B(2)_i^\pm,\ldots$ apart from a
finite number of them.  Since
$U(B_i^\pm;\Phi_1,\ldots,\Phi_s;\epsilon)=\cap_{k=1}^s
U(B_i^\pm;\Phi_k;\epsilon)$, it is sufficient to show that for any
neighborhood $U(B_i^\pm;\Phi;\epsilon)$ there exists an integer $N$
such that $B(p)_i^\pm \in U(B_i^\pm;\Phi;\epsilon)$ for any $p> N$ or,
which is the same, see~(\ref{e5-20}), that
\beq
|(B(p)_i^\pm  -  B_i^\pm) \Phi|_0 < \epsilon, \q\forall~p>N. 
\label{e5-22}
\eeq
The above equation has to hold for any $\Phi$ and any $\epsilon$.
In general $N$ depends on $\Phi$ and $\epsilon$, $N=N(\Phi,\epsilon)$.

The fact that $B_i^+ - B(p)_i^+$ is a continuous linear
operator in $\Omega$ is essential since relations like~(\ref{e5-18}) can
be used. The latter together with~(\ref{e5-11a})-(\ref{e5-11b})
and~(\ref{e5-14}) yields~:
\bea
(B_i^+ - B(p)_i^+)\Phi &=& \sum_{l_1+\cdots+l_n <
p}c(L)(\sqrt{l_i+1} \Biggl(1-\sqrt{1-{{\sum_{k} l_k}\over p
}}\Biggr)|L\rangle_i \nn\\
&&+\sum_{l_1+\cdots+l_n\ge
p}c(L)(\sqrt{l_i+1}|L\rangle_i.
\label{e5-23}
\eea
The continuity of the scalar product with respect to the
$cn$-topology and the fact that all terms in the RHS of~(\ref{e5-23}) 
are orthogonal to each other yield~:
\beas
(|(B_i^+ - B(p)_i^+)\Phi|_0)^2 &=&
\sum_{l_1+\cdots+l_n< p}|c(L)|^2 (l_i+1)
\Biggl(1-\sqrt{1-{{\sum_{k} l_k}\over p }}\Biggr)^2 \\
&&+\sum_{l_1+\cdots+l_n\ge p}|c(L)|^2(l_i+1).  
\eeas
Let $\epsilon >0$. Select $p_0\in \N$ to be fixed. For any
$p>p_0$
\bea
&&(|(B_i^+ - B(p)_i^+)\Phi|_0)^2 =
  \sum_{l_1+\cdots+l_n\le p_0}|c(L)|^2 (l_i+1)
  \Biggl(1-\sqrt{1-{{\sum_{k} l_k}\over p }}\Biggr)^2 \nn\\
&& +\sum_{p_0<l_1+\cdots+l_n < p}|c(L)|^2 (l_i+1)
  \Biggl(1-\sqrt{1-{{\sum_{k} l_k}\over p }}\Biggr)^2
+ \sum_{l_1+\cdots+l_n\ge p}|(1+l_i)c(L)|^2  \nn \\
&& < \sum_{l_1+\cdots+l_n\le p_0}|c(L)|^2 (l_i+1)
  \Biggl(1-\sqrt{1-{{\sum_{k} l_k}\over p }}\Biggr)^2
+ \sum_{l_1+\cdots+l_n>p_0}|(1+l_i)c(L)|^2.   \label{e5-24}
\eea
Since the partial sums of
$\sum_{l_1,\ldots,l_n=0}^\infty(1+l_i)|c(L)|^2$
constitute an increasing sequence of positive numbers, 
which is restricted from
above, $\sum_{l_1,\ldots,l_n=0}^\infty(1+l_i)|c(L)|^2\le|\Phi|_1$,
the series $\sum_{l_1,\ldots,l_n=0}^\infty(1+l_i)|c(L)|^2$ converges.
Choose $p_0$ such that
$\sum_{l_1+\ldots+l_n>p_0}(1+l_i)|c(L)|^2<{{\epsilon^2}\over 2}$.
Then for any $p>p_0$
\bea
&& (|(B_i^+ - B(p)_i^+)\Phi|_0)^2 <
  \sum_{l_1+\cdots+l_n\le p_0}|c(L)|^2 (l_i+1)
  \Biggl(1-\sqrt{1-{{\sum_{k} l_k}\over p }}\Biggr)^2 +
  {{\epsilon^2}\over 2}  \nn\\
&& < \sum_{l_1+\cdots+l_n\le p_0}|c(L)|^2 (l_i+1)
    \Biggl(1-\sqrt{1-{{p_0}\over p }}\Biggr)^2 +
  {{\epsilon^2}\over 2}
  <  d \Biggl(1-\sqrt{1-{{p_0}\over p }}\Biggr)^2 +
  {{\epsilon^2}\over 2},
\label{e5-25}
\eea
where
$d=\sum_{l_1+\cdots+l_n\le p_0}|c(L)|^2 (l_i+1)$ is a constant.
Clearly there exists $N\in \N$ such that
$d \Bigl(1-\sqrt{1-{{p_0}\over p }}\Bigr)^2 <
 {{\epsilon^2}\over 2}$ for any $p>N$.
Hence for every $\epsilon > 0$ there exists a positive integer
$N$ such that $|(B_i^+ - B(p)_i^+)\Phi|_0 < \epsilon$, $\forall
p> N$, i.e.\ (\ref{e5-22}) holds.

In a similar way one proves that $\displaystyle\lim_{p\to \infty}
B(p)_i^- =B_i^- $. This completes the proof. \mybox


\section{Bosonization of $A$-statistics}
\setcounter{equation}{0}

A simple comparison of~(\ref{e3-19a})-(\ref{e3-19b}) 
with~(\ref{e5-14}) suggests
that the Jacobson CAOs of any order $p$ can be bosonized, namely
that they can be expressed as functions of Bose CAOs
$B_1^\pm,\ldots,B_n^\pm$, see~(\ref{e5-15}). Indeed, taking into account
that $B_i^+B_i^-\equiv N_i$ is a number operator for bosons in a
state $i$,
\beq
N_i|L\rangle\equiv
N_i|l_1,\ldots,l_i,\ldots,l_n\rangle=
l_i|l_1,\ldots,l_i,\ldots,l_n\rangle,\q i=1,\ldots,n,
\label{e6-1}
\eeq
one rewrites~(\ref{e3-19a}) as~:
\[
a_i^+|L\rangle=
  \sqrt{(l_i+1)(p-{\sum_{k=1}^n N_k+1})}~
|L\rangle_{i}.
\]
In view of~(\ref{e5-14}) the latter can also be represented as
\beq
a_i^+|L\rangle=
  \sqrt{p+1-\sum_{k=1}^n N_k}~B_i^+|L\rangle=
B_i ^+\sqrt{p-\sum_{k=1}^n B_k^+B_k^-}~|L\rangle.
\label{e6-2}
\eeq
Since~(\ref{e6-2}) holds for any $|L\rangle$,
\beq
a_i^+= B_i ^+\sqrt{p-\sum_{k=1}^n B_k^+B_k^-}, \quad
 i=1,\ldots,n.
\label{e6-3}
\eeq
In a similar way one derives from~(\ref{e3-19b})~:
\beq
a_i^- =\sqrt{p-\sum_{k=1}^n B_k^+B_k^-}~B_i^-,\quad
 i=1,\ldots,n.  
\label{e6-4}
\eeq
Evidently also, see~(\ref{e3-20a}),
\beq
h_0=p-\sum_{k=1}^n B_k^+B_k^-. 
\label{e6-5}
\eeq
Note that the entire Fock space $W$ is reducible with respect to
the Jacobson CAOs. Its finite-dimensional ``physical" subspace
$W_p$, see~(\ref{e5-10}), is a simple (= irreducible) $gl(n+1)$-module
and within this module $(a_i^+)^*=a_i^- $ holds.

After simple calculations and taking into account that
$a_i^+=e_{i0}$, $a_i^-=e_{0i}$, $i=1,\ldots,n$, see~(\ref{e2-3}), one can
express all Weyl generators $\{e_{ij}|i,j=0,1,\ldots,n\}$ of
$gl(n+1)$ via $n$ pairs of Bose operators~:
\bea
(a)&& e_{ij}=B_i^+B_j^-, \quad i,j=1,\ldots,n, \nn\\
(b)&& e_{i0}=B_i^+\sqrt{p-\sum_{k=1}^n B_k^+B_k^-}, \quad
e_{0i}=\sqrt{p-\sum_{k=1}^n B_k^+B_k^-}~B_i^-, \quad
i=1,\ldots,n, \label{e6-6} \\
(c)&& e_{00}=p-\sum_{k=1}^n B_k^+B_k^-, \nn
\eea
where, we recall, $p$ is any positive integer, $p\in\N$.

The above bosonization of $gl(n+1)$ is not unknown.  Up to a
choice of notation it is the same as the so-called
Holstein-Primakoff (H-P) realization of $gl(n+1)$~\cite{ok75}, 
initially introduced for $sl(2)$~\cite{ho40, dy56}.  
Note that~(\ref{e6-6}a) alone gives
the known Jordan-Schwinger realization of $gl(n)$ via $n$ pairs
of Bose operators.

\section{Other applications~: a two-leg $S=1/2$ quantum Heisenberg
ladder}
\setcounter{equation}{0}

In the present section we show that the Jacobson CAOs may also be 
of more general interest. We demonstrate this on the example of a
two-leg $S=1/2$ Heisenberg spin ladder~\cite{go94, su98}, where the
Jacobson CAOs of order $p=1$ appear in a natural way. The
considerations below hold however for several other Heisenberg
spin models (examples include lattice models with 
dimerization~\cite{sa90, ch91b, ch91}, 
two-layer Heisenberg models~\cite{ch95, ko98, sh00}) and
more generally for any hard-core Bose model~\cite{fi89} with degenerated
orbitals per site (as for instance in~\cite{zi94, la98}).

The Hamiltonian of the model reads:
\beq
{\hat H}=\sum_{i} (J{\bf {\hat S}}_{i}^+ {\bf {\hat S}}_{i+1}^+ +
J{\bf {\hat S}}_{i}^- {\bf {\hat S}_{i+1}}^- + J_\bot{\bf {\hat
S}}_{i}^+ {\bf {\hat S}}_{i}^-). 
\label{e7-1}
\eeq
Here
${\bf {\hat S}}_{i}^\pm\equiv ({\hat S}_{1 i}^\pm,{\hat S}_{2 i}^\pm,
{\hat S}_{3 i}^\pm)$ are two commuting
spin-$1/2$ vector operators ``sitting" on site $i$ of the chain
$\pm$ and the Hamiltonian is a scalar with respect to the total
spin operator ${\bf {\hat S}}
=\sum_{i}( {\bf {\hat S}}_{i}^+ + {\bf {\hat S}}_{i}^-)$~:
\beq
[{\hat S}_{\alpha i}^\pm,{\hat S}_{\beta i}^\pm]=
i\sum_{\gamma}\epsilon_{\alpha \beta \gamma}{\hat S}_{\gamma
i}^\pm,\q 
[{\hat S}_{\alpha i}^+,{\hat S}_{\beta j}^-]=0, \q
[{\hat H},{\bf {\hat S}}]=0. 
\label{e7-2}
\eeq

Every local state space $W_i$ related to site $i$ is
4-dimensional with a basis $|\uparrow, \uparrow \rangle$, $|\uparrow,
\downarrow \rangle$, $|\downarrow, \uparrow \rangle$, $|\downarrow,
\downarrow \rangle$ and $W=W_1\otimes W_2 \otimes \ldots \otimes W_N$ is
the global state space of the system (in the case of a ladder with
$N$ sites). The notation is standard~: if $A$ is any operator in
$W_i$, then the corresponding to it operator in $W$ is denoted as
$A_i$, where $A_i\equiv id_1\otimes\ldots \otimes id_{i-1}\otimes A
\otimes id_{i+1} \otimes \ldots \otimes id_N$.

If the system is in a disordered phase ( $J_\bot \gg J$)
its state is well described with the bond operator representation
of spin operators~\cite{ch89, sa90}, which is a particular kind of
bosonization~:
\beq
{\hat S}_{\alpha i}^\pm={1\over 2}(\pm B_{\alpha i}^- \pm
B_{\alpha i}^+ -i\epsilon_{\alpha \beta \gamma}B_{\beta
i}^+B_{\gamma i}^-),\q
\alpha, \beta, \gamma=1,2,3. 
\label{e7-3}
\eeq
Here $B_{1i}^\pm$, $B_{2i}^\pm$, $B_{3i}^\pm$ are three pairs of
Bose CAOs related to  site $i$ and the vectors
$|0\rangle_i$, $B_{1i}^+|0\rangle_i$, $B_{2i}^+|0\rangle_i$, 
$B_{3i}^+|0\rangle_i$
constitute another basis in $W_i$.

The treatment of the model in terms of bosonic  operators is
advantageous because of the simpler commutation rules of Bose
statistics. It rises however certain problems. As mentioned above,
any local state space $W_i$ is 4-dimensional, whereas the local Bose
Fock space $\Phi_i$ is infinite-dimensional. Moreover $W_i$ is not
invariant in $\Phi_i$ with respect to the Bose CAOs (and, as a
result, with respect to the local spin operators~ (\ref{e7-3})). The
physical state space $W$ is not an invariant subspace of the
global Fock space $\Phi=\Phi_1\otimes \Phi_2 \otimes \ldots \otimes
\Phi_N$ with respect to the Hamiltonian~(\ref{e7-1}).

Various approaches have been proposed in order to overcome the
problem. Following~\cite{sa90},
additional scalar bosons $s_i^\pm$ were
introduced in~\cite{go94}. 
Then the physical states are those which satisfy an
additional constraint $s_i^+s_i+ \sum_\alpha
B_{i\alpha}^+B_{i\alpha}=1$.  Another way is to keep the
realization~(\ref{e7-3}) but to introduce ``by hands" a fictitious
infinite on-site repulsion between the ``bosons"~\cite{ko98} (first
proposed in~\cite{fi89} for a nondegenerate case). This forbids
configurations with two or more bosons accommodated on one and the
same site.  The latter leads to the ``hard-core" condition
$B_{\alpha i}^\pm B_{\beta i}^\pm=0$, i.e.\ the hard-core bosons
are not quite bosons, since they satisfy fermionic-like
conditions.

A third approach was worked out in~\cite{ch89} 
(see also~\cite{ch91b, ch91, ch95}).
It proposes the Bose operators $B_{\alpha i}^\pm$ in~(\ref{e7-3}) to be
replaced throughout by new operators $b_{\alpha i}^\pm$ as
follows~:
\beq
B_{\alpha i}^+ ~ \rightarrow ~ b_{\alpha i}^+= B_{\alpha i}^+
\sqrt{1-\sum_{\beta=1}^3 B_{\beta i}^+B_{\beta i}^-},\q
B_{\alpha i}^- ~ \rightarrow ~ b_{\alpha i}^-=
\sqrt{1-\sum_{\beta=1}^3 B_{\beta i}^+B_{\beta i}^-}~B_{\alpha
i}^-.
\label{e7-4}
\eeq
A simple comparison with~(\ref{e6-3}), (\ref{e6-4}) indicates that
\begin{itemize}
\item
The Bose operators related to site $i$, i.e.\
$B_{1i}^\pm$, $B_{2i}^\pm$, $B_{3i}^\pm$, are replaced by $p=1$
Jacobson CAOs (or, which is the same, by $p=1$ quasi-Bose
operators),
\beq
B(1)_{\alpha i}^\pm \equiv b_{\alpha i}^\pm,\q\alpha=1,2,3,
\label{e7-5}
\eeq
in their Holstein-Primakov realization. Consequently 
(Proposition~\ref{prop3}) the hard-core condition 
$b_{\alpha i}^+b_{\beta i}^+=0$ holds;
\item
The Jacobson CAOs from different sites commute~:
\beq
[b_{\alpha i}^\xi, b_{\beta j}^\eta]=0,\q \hbox{if }
i\ne j\hbox{ for any }\xi,\eta=\pm\hbox{ and } \alpha, \beta=1,2,3.
\label{e7-6}
\eeq
\end{itemize}
It is essential that the substitution~(\ref{e7-4}) does not change the
commutation relations~(\ref{e7-2}) between the new spin operators
\beq
S_{\alpha i}^\pm={1\over 2}(\pm b_{\alpha i}^- \pm b_{\alpha i}^+
-i\epsilon_{\alpha \beta \gamma}b_{\beta i}^+b_{\gamma i}^-),\q
\alpha, \beta, \gamma=1,2,3,
\label{e7-7}
\eeq
and the corresponding new Hamiltonian
\beq
H=\sum_{i} (J{\bf S}_{i}^+ {\bf {S}}_{i+1}^+ + J{\bf { S}}_{i}^-
{\bf { S}_{i+1}}^- + J_\bot{\bf { S}}_{i}^+ {\bf { S}}_{i}^-).
\label{e7-8}
\eeq
Moreover each local state space $W_i$ is an invariant subspace of
$\Phi_i$ with respect to the Jacobson CAOs and hence with respect
to any function of them (in particular with respect to the spin
operators~(\ref{e7-7})).  
The Hamiltonian~(\ref{e7-8}) is also a well defined
operator in $W$.

The conclusion is that replacing throughout the model the Bose
operators with $p=1$ Jacobson CAOs $b_{\alpha i}^\pm$, which
commute at different sites (see~(\ref{e7-6})), one obtains directly the
physical state space and the correct expressions for the spin
operators and the Hamiltonian. There is no need to introduce
either a fictitious infinite-dimensional repulsion or
additional relations.  All these requirements are already encoded
in the properties of the Jacobson CAOs.

Let us point out that the above results can be also derived from
the following proposition, which is of independent interest.

\begin{prop}
Let $B_\alpha^\pm$, $\alpha=1,\ldots,n$,
be $n$ pairs of Bose CAOs with a Fock space $\cal F$ and a 
basis~(\ref{e5-6}). 
Denote by ${\cal F}_1$ the subspace of $\cal F$ linearly
spanned on the vacuum and all ``single-particle" states,
\beq
{\cal F}_1=\hbox{span}\{|l_1,\ldots,l_n\rangle |~ 
l_1+\cdots+l_n\le 1\}.
\label{e7-9}
\eeq
Let $\cal P$ be a projection operator of $\cal F$ onto ${\cal F}_1$~:
\beq
{\cal P}|l_1,\ldots,l_n\rangle =
\left\{
\begin{array}{ll}
|l_1,\ldots,l_n\rangle, &\hbox{ if } l_1+\cdots+l_n\le 1;\\
0, & \hbox{ if } l_1+\cdots+l_n> 1.
\end{array} \right.
\label{e7-10}
\eeq
Then the operators ${\cal P}B_\alpha^\pm{\cal P}$,
$\alpha=1,\ldots,n$, considered as operators in ${\cal F}_1$, are
$p=1$ Jacobson CAOs,
\beq
{\cal P}B_\alpha^\pm{\cal P}=B(1)_\alpha^\pm\equiv b_\alpha^\pm, \q
\alpha=1,\ldots,n.
\label{e7-11}
\eeq
\label{prop5}
\end{prop}

\noindent {\it Proof.} 
One verifies directly that (\ref{e2-5}) and (\ref{e3-22})
hold. \mybox

Coming back to the two-leg spin ladder model, introduce a
projection operator ${\cal P}_w={\cal P}_1\otimes{\cal
P}_2\otimes\ldots\otimes {\cal P}_N$ of $\Phi$ onto $W$, where each
${\cal P}_i$ projects $\Phi_i$ onto $W_i$ according to~(\ref{e7-10})
with $n=3$. The projector ${\cal P}_w$ provides an alternative way
for writing down the expressions for the spin operators~(\ref{e7-7}) and
the Hamiltonian~(\ref{e7-8}). 
Instead of using the substitution~(\ref{e7-4}),
one can set:
\beq
H={\cal P}_w {\hat H} {\cal P}_w,~~ S_{\alpha i}^\pm={\cal
P}_i{\hat S}_{\alpha i}^\pm {\cal P}_i,\q
i=1,\ldots,N.
\label{e7-12}
\eeq
The operator ${\cal P}_w$ is a Bose analogue of the Gutzwiller
projection operators~\cite{gu63}, 
extensively used in the $t$-$J$ models in
order to exclude the double occupation of fermions at each site
(see, for instance~\cite{ts94} where a similar problem, a $t$-$J$ two-leg
ladder is investigated).

\section{Concluding remarks}
\setcounter{equation}{0}

{}From a mathematical point of view the JGs 
$a_1^\pm,\ldots,a_n^\pm$ provide a new description 
of the Lie algebra $sl(n+1)$
in terms of generators and relations~(\ref{e2-5}),
based on the concept of Lie triple systems.
For the same reason any $n$ pairs of parafermions (resp.\
parabosons) can be called Jacobson generators of the orthogonal Lie
algebra $so(2n+1)$ (resp.\ of the orthosymplectic Lie
superalgebra $osp(1/2n)$). The JGs provide an alternative to
the Chevalley descriptions of these Lie (super)algebras. 

{}From a physical point of view the interest 
in the JGs of $sl(n+1)$ stems from the observation 
that they indicate the possible existence of 
a new quantum statistics.
Indeed, we have seen that within each Fock space $W_p$ the
operator $a_i^+$ (resp.\ $a_i^-$) can be interpreted as
an operator creating (resp.\ annihilating) a particle, a quasiboson
in a state $i$ (in particular with an energy  $\e_i$). 

In many respects the quasibosons behave as bosons.
Similar as for bosons, the quasibosons
can be distributed along the orbitals in an
arbitrary way as far as the number of
accommodated particles $M$ does not exceed $p$.
The number of different states of $M\le p$ quasibosons
is the same as for bosons (the $M$-particle subspaces
of quasibosons and bosons have one and the same dimension). 
There is however one essential difference~:  
quasiboson systems of order $p$ can accommodate at most 
$p$ particles. 

In order to use a proper Lie algebraic language
we have restricted our
considerations to finite-dimensional Lie algebras.
In other words, we were studying systems with a finite
number $n$ of orbitals.
Such systems certainly do exist. Examples are the local state
spaces of spin systems (in particular the example
considered in Section~7), $su(n)$ lattice models etc.
Nevertheless it is natural to ask whether 
$A$-statistics can be extended to incorporate
infinitely many orbitals   
as this is usual in quantum theory.
The answer to this question is positive and it
is in fact evident from the results we have obtained 
so far. First of all the description of $sl(n+1)$
via generators~(\ref{e2-3}) and relations~(\ref{e2-5}) is well defined
for $n=\infty$, namely for $sl(\infty)$. Secondly, any
Fock module $W_p$ as given in Corollary~\ref{cor2} and in particular
equations~(\ref{e3-9}) are also well defined  
for $n=\infty$. In this case any $W_p$ is an irreducible
$sl(\infty)$ module, generated out of the vacuum
by means of the Jacobson creation operators.
Therefore each state $|p;l_1,\ldots,l_i,\ldots\rangle$
contains no more than a finite number of nonzero entries $l_i$.
Moreover due to Proposition~\ref{prop3}
the physical state space is a linear span of all vectors
$|p;l_1,\ldots,l_i,\ldots\rangle$ with
\beq
l_1+l_2+\cdots+l_i+\cdots \le p. 
\label{e8-1}
\eeq
All such states constitute an (orthonormal) basis in 
$W_p$. They transform according to the same 
relations~(\ref{e3-19a})-(\ref{e3-19b}) with $n=\infty$.
It is straightforward to verify that any 
$sl(\infty)$ module $W_p$  is a Fock space in the sense of
Definition~\ref{def1}. Finally, the Pauli principle 
(Corollary~\ref{cor3})
remains valid also for $n=\infty$~: despite of the
infinitely many available orbitals, the infinitely many places
to be occupied by the quasibosons, the system
cannot accommodate more than $p$ particles.

The indices labelling the CAOs in QFT do not constitute
a countable set. For instance the Bose CAOs $a({\bf p})^\pm$
of a scalar field are labelled by the momentum
of the particle, which takes values in
$\R^3$.   
Also in this case the above considerations remain valid, but now, as
in the canonical Bose case, the CAOs are operator valued distributions
(and $\delta_{ij}$ is $\delta(i-j)$). 
More generally, the indices labelling the CAOs can take values 
in spaces of any dimension.   
Therefore $A$-statistics 
is not restricted to $1D$ or $2D$ spaces only.  

We should point out that within $A$-statistics the main
quantization equation~(\ref{e2-6}) does not determine uniquely the
creation and annihilation operators. The Jacobson generators~(\ref{e2-3})
yield one possible solution of~(\ref{e2-6}). For another possible choice
(a causal $A$-statistics), we refer to~\cite{pa79}.

The quasi-Bose operators $B(p)_1^\pm,\ldots, B(p)_n^\pm$,
introduced in Section~5 can be used as an
approximation, in fact a good approximation, to Bose statistics
for values of the order of statistics $p$, which is much bigger
than the number of accommodated particles. An additional advantage
of the quasi-Bose CAOs of any order $p$ is that they are bounded
linear operators, defined everywhere in the Fock space $W_p$.
This property avoids the rather delicate questions of whether
the operators under consideration can be defined on a common
domain of definition $\Omega$, so that any polynomial of them  is
also well defined in $\Omega$.

The ``opposite" to $p \rightarrow \infty$ case, namely the $p=1$
Jacobson CAOs (or, which is the same, the $p=1$ quasi-Bose
operators) turns out to be of interest too. We have illustrated this on
a particular example from condensed matter physics. 
The $p=1$ quasiboson representation appears naturally 
in lattice Bose models with infinitely strong repulsion between 
the particles, which forbids configurations with
more than one particle per site.
One can speculate that representations with order
of statistics $p$ could be of interest in pictures where
no more that $p$ particles can be accommodated on each
site of the lattice.

For applications of quasiboson representations in nuclear theory
we refer to~\cite{me98}. 
As indicated there, the $p=1$ quasi-Bose operators 
reduce to Klein-Marshalek algebras~\cite{kl88}, 
which are extensively used in
nuclear physics.

One way to enlarge the class of statistics studied here is
to deform the relations~(\ref{e2-5}) or, which is the same, to
deform $sl(n+1)$ so that the main quantization equation~(\ref{e2-6})
remains unaltered. 
The possibility for such deformations stems from the 
observation that the commutation relations between the Cartan 
elements (the Hamiltonian is a Cartan element, see~(\ref{e2-18}))
and the root vectors (the Jacobson generators are root vectors,
see~(\ref{e2-4})) remain unaltered upon quantum deformations 
($q$-deformations). Therefore the problem actually is to
express the known $q$-deformations of $sl(n+1)$ via deformed
Jacobson generators. This is the first step. The second
step will be to define the Fock representations and to write
down the deformed analogue of~(\ref{e3-19a})-(\ref{e3-19b}). 
Partial results
in this respect were already announced~\cite{par98,pa00}.

\section*{Acknowledgments}
One of us (T.D. Palev) is grateful to Prof.\ Randjbar-Daemi for the 
kind hospitality at
the High Energy Section of ICTP. Constructive discussions with
Dr.\ N.I.\ Stoilova are greatly acknowledged.  This work was
supported by the Grant $\Phi$-910 of the Bulgarian Foundation for
Scientific Research, and by Grant PST.CLG.976865 of NATO.

\end{document}